\documentclass{aa}  
\usepackage{bm}
\usepackage{graphicx}
\usepackage{txfonts}
\usepackage[colorlinks=true,linkcolor=blue,citecolor=blue,urlcolor=blue]{hyperref}
\usepackage[dvipsnames]{xcolor}
\usepackage[normalem]{ulem}
\usepackage{soul}

\newcommand{\yr}{$\mathrm{yr}$}

\newcommand{\mdot}{$\dot{M}$}
\newcommand{\Msun}{$\,M_{\odot}$}

\newcommand{\kms}{$\,~\textrm{km\,s}^{-1}$}

\usepackage{xcolor}
\usepackage{ifthen}

\usepackage{etoolbox}

\newbool{showred}
\setbool{showred}{false} 

\newcommand{\new}[1]{%
  \ifbool{showred}
    {\textcolor{red}{#1}}
    {#1}%
}

\begin{document} 

\authorrunning{
    Larkin et al.}

    \title{Investigating dusty red supergiant outflows in Westerlund 1 with 3D hydrodynamic simulations}

   \author{C. J. K. Larkin\inst{\ref{inst:mpik},\ref{inst:ari},\ref{inst:mpia}}
          \and
          J. Mackey\inst{\ref{inst:dias}}
          \and
          T. J. Haworth\inst{\ref{inst:qmul}}
           \and
          A. A. C. Sander\inst{\ref{inst:ari}}
          }

   \institute{{Max-Planck-Institut f\"{u}r Kernphysik, Saupfercheckweg 1, D-69117 Heidelberg, Germany\label{inst:mpik}}\\
            \email{cormac.larkin@mpi-hd.mpg.de}
         \and 
         {Astronomisches Rechen-Institut, Zentrum f\"{u}r Astronomie der Universit\"{a}t Heidelberg, M\"{o}nchhofstr. 12-14, D-69120 Heidelberg, Germany\label{inst:ari}}
         \and 
         {Max-Planck-Institut f\"{u}r Astronomie, K\"{o}nigstuhl 17, D-69117 Heidelberg, Germany\label{inst:mpia}}
         \and 
          {Astronomy \& Astrophysics Section, School of Cosmic Physics, Dublin Institute for Advanced Studies, DIAS Dunsink Observatory, Dublin D15 XR2R, Ireland\label{inst:dias}}
         \and 
         {Astronomy Unit, School of Physics and Astronomy, Queen Mary University of London, London E1 4NS, UK\label{inst:qmul}}
             }

   \date{Received DATE; accepted DATE}

  \abstract
   {Recent JWST observations towards Westerlund 1 have revealed extensive nebular emission associated with the cluster. Given the age of the region and  the proximity of that material to massive stars, it cannot be primordial star-forming gas and the origin is uncertain. } 
   {We aim to determine whether the nebular emission in Westerlund 1 is due to ablation flows from red supergiant (RSG) stars embedded in the cluster wind driven by the Wolf-Rayet stars in the cluster core. We also aim to explore the efficiency of mass loading for the RSG wind in this scenario.}
   {We used 3D hydrodynamic simulations with the \textsc{pion} code to study the interaction between the cluster and RSG winds. We compared our simulations with the JWST observations by generating synthetic dust-emission maps.}
   {We find that the ablation flow morphology -- which shows clumps and instabilities -- is consistent with the observations towards Westerlund 1. Synthetic observations at 11\,$\mu$m predict fluxes in the ablation flow of $\sim1000-6000$\,MJy\,ster$^{-1}$, which is consistent with the unsaturated components of the JWST F1130W observations in the vicinity of the RSGs in the region. This good agreement is achieved without any consideration of polycyclic aromatic hydrocarbons (PAHs), which have a known 11.3\,$\mu$m feature that appears in the F1130W band. This suggests that the environment is not conducive to PAH formation and/or the ablation flow is PAH-depleted by wind and radiation action.}
   {Ablation of RSG winds can explain the observed nebulosity in Westerlund 1, at least in the vicinity of the RSGs. Further observations are encouraged to enable detailed studies of these interactions.}


   \maketitle
\nolinenumbers

\section{Introduction}

Young massive stellar clusters (YMCs) provide ideal environments to study the evolution of massive stars, planet and star formation processes, and stellar feedback to the Galaxy \citep{PortegiesZwart2010}. In recent times, they have also been proposed as sources of cosmic rays due to detections of gamma rays \citep[e.g.][]{2022A&A...666A.124A}. Westerlund 1 is thought to be the most massive YMC in the Galaxy, with mass estimates of the order of $5\times 10^4$ to $10^5 M_{\odot}$ \citep{ClaNegCro05, AndGenBra17}, although the inferred mass is dependent on the derived luminosity (and hence the age) of the massive stars in the cluster, which is quite uncertain because of the large extinction to the cluster \citep{BeaDavSmi21}. 
Westerlund 1 hosts a large and diverse population of massive stars at various evolutionary stages, with 24 Wolf-Rayet (WR) stars \citep{2006MNRAS.372.1407C} as well as OB supergiants, yellow supergiants, red supergiants (RSGs), luminous blue variables, and an sgB[e] star \citep{2020A&A...635A.187C}. The WR stars drive a cluster wind of several thousand km\,s$^{-1}$ \citep{Haerer2023} that sculpts a superbubble cavity around the cluster. The region around Westerlund 1 is a bright gamma-ray emitter to very high energies \citep{2022A&A...666A.124A}, likely due to the combined action of stellar winds and supernovae in accelerating cosmic rays \citep[e.g.][]{Vieu2023,Haerer2023}.

The concept of mass loading of a fast wind by hydrodynamic (HD) mixing with cold and slow material was introduced by \citet{1986MNRAS.221..715H}, who applied it to the expansion of WR nebulae and the consequent X-ray and nebular emission that should arise according to their model. Mass loading in stellar clusters was considered theoretically by \cite{Stevens2003} in the context of protostars and protoplanetary disks and their effects on diffuse X-ray emission. The effects of stellar winds and supernovae on molecular material remaining from the formation of a YMC were modelled by \cite{Rogers2013}, who found the cluster wind to be mass loaded by a factor of up to several hundred. \cite{Arthur2012} modelled the X-ray emission from the mass loading of protoplanetary disks in Orion. Mass loading of the cluster wind by the dense and slow winds of cool supergiants reduces the thermal pressure driving the cluster wind \citep{2011ApJ...743..120S}, potentially limiting its effectiveness in creating a superbubble and accelerating cosmic rays.

Observations of diffuse thermal X-ray emission in Westerlund 1 show a hard component \citep{Kavanagh2011,Haubner2025}. While pre-main-sequence stars are thought to contribute to this \citep{Clark2008}, the majority is attributed to thermalisation of the winds from WR stars in the cluster core. These winds must mix with cooler material to explain the gas temperature derived from X-ray observations, but the origin of this material is not clear. Dust and ejecta from supernovae are thought to not contribute significantly to this process  \citep{Kavanagh2011,Haubner2025} since the supernova rate of 1 per $\sim$$10\,$kyr would imply the cluster is free of supernova remnant material for $\sim$$90\%$ of the time \citep{Muno2006}; however, we note that this estimated supernova rate is very uncertain. RSG winds inject cool material that can be ionised by an external radiation field, such as that found in a YMC \citep{Mackey2014}. This has already been demonstrated for W26, one of the four RSGs in Westerlund 1 \citep{2014MNRAS.437L...1W, Mackey2015}. Furthermore, the potential for the enriched RSG wind material to contribute to future star formation within the cluster depends on whether it is heated and ejected rapidly, or can remain confined, cool, and dense \citep{Mackey2015}.

Recent JWST images of Westerlund 1 from the Extended Westerlund 1 and 2 Open Clusters Survey (EWOCS) show resolved shells and outflows from the four RSGs and other evolved cool massive stars in the cluster \citep{Guarcello2024}. The emission mechanism for gas and dust in Westerlund 1 is not clear, but its proximity to the core of such a massive cluster implies that the material was recently ejected from the cluster member stars. If the mid-IR emission is produced by dust, then the most likely source is winds and mass ejections from the cool supergiants and hypergiants in the cluster core.

In this paper we explore, in a sense, the opposite case to that of \cite{Rogers2013}, that of a cool RSG wind embedded in a hot, fast collective cluster wind from a YMC. We present a 3D HD simulation of our scenario to explore the efficiency of mass loading from evolved stellar winds, and use it to create synthetic dust-emission maps. 
There is some similarity between our work and previous simulations of the G2 cloud orbiting the Galactic Centre: \cite{2014ApJ...789L..33D} and \cite{Ballone2018} modelled a cool star emitting a slow wind and moving rapidly through the hot coronal gas around the supermassive black hole Sgr A$^\ast$. To our knowledge, no multi-dimensional simulations have been made of the interaction of stellar wind from a RSG embedded in a hot star-cluster wind, though such a situation must be common in YMCs.

Our paper is organised as follows: In Sect. \ref{sec:methods} we discuss our simulation setup and the values adopted for the RSG and cluster winds. In Sect. \ref{sec:results} we present the results of our simulation, including slices through the XY domain (Sect. \ref{sec:results_slices}), phase diagrams for the RSG wind and cluster wind material (Sect. \ref{sec:results_phase}), mass loading of the cluster wind over time (Sect. \ref{sec:results_ML}), synthetic dust continuum-emission maps for 11\,$\mu$m and 24\,$\mu$m derived from our simulation (Sect. \ref{sec:results_dust}), and how those maps correspond to observations from JWST (Sect. \ref{sec:results_comp}). In Sect. \ref{sec:discussion} we discuss the uncertainties in the winds of RSGs and clusters (Sect. \ref{sec:discussion_uncert_wind}) and in our modelling (Sect. \ref{sec:discussion_uncert_model}), as well as the implications for this work in understanding the role of mass loading (Sect. \ref{sec:discussion_ML}), its connection to super star clusters (Sect. \ref{sec:discussion_SSC}), and the need for follow-up observations (Sect. \ref{sec:discussion_obs}). We present our conclusions in Sect. \ref{sec:conclusion}, and a resolution study in Appendix \ref{app:res_test}.

\section{Methods}
\label{sec:methods}

\subsection{HD simulations}
We performed our 3D Cartesian simulations using the (magneto-)HD code \textsc{pion} \citep{2021PION}. We employed the static mesh refinement option in \textsc{pion} to achieve high resolution only where required. 
For our radiative heating and cooling prescription, we used model 8 in \textsc{pion} as described in \citet{Green2019}, which assumes a sufficiently intense extreme-ultraviolet (EUV) radiation field to keep H and He photoionised everywhere on the domain, and which includes three cooling sources. The first source is the maximum of the Solar metallicity collisional ionisation equilibrium curve described by \cite{Wiersma2009}, and the cooling function for forbidden lines described by \cite{Henney2009}. This ensures we account for the significant rates of cooling that occur in photoionisation equilibrium at $T\sim10^4$\,K. The second source included is ionised hydrogen \citep{Hummer1994} and helium \citep{Rybicki1979} Bremsstrahlung. The third source is the cooling rate of hydrogen recombination from \cite{Hummer1994}, where we assumed hydrogen is fully ionised. 
Heating is calculated by assuming each H recombination (excluding recombinations directly to the ground state, i.e. the Case B approximation) rapidly results in a re-ionisation by the intense Lyman-continuum radiation field found in YMCs, with associated photo-heating of 5\,eV per photoionisation.
We obtain an equilibrium gas temperature $T\approx8300$\,K for this prescription.

In Westerlund 1 the EUV radiation field is provided by the OB and WR stars in the cluster, which are observed to ionise the circumstellar nebulae of the cool stars to a high degree in the vicinity of the cluster \citep{2018MNRAS.477L..55A}. 
This validates our above assumption that the medium is photoionised.
Consistent with this observation, postprocessing of simulation snapshots in Sect.~\ref{sec:results_dust} with the \textsc{torus} radiation transfer code calculates the ionisation state of the gas as well as dust emission.  From this we find that H is more than 99\% ionised in the full domain except for within a radius of $\approx3\times10^{16}$\,cm around the RSG where the wind is dense enough to self-shield from ionising radiation.

The simplified microphysics treatment is quite standard in HD modelling of nebulae \citep[e.g.][]{GarciaSegura1996a, Meyer2014, Mackey2025}, where calculation of radiative transfer and non-equilibrium ionisation would make the simulations prohibitively expensive \citep{Mathew2025}.
\citet{Haworth2015} investigated the differences between a simplified microphysics treatment and more complicated calculations with full radiative transfer and ionisation of H, He, and metals, applied to expanding H~\textsc{ii} regions.
They found that differences were predominantly driven by the equilibrium temperature of the photoionised gas, determined by the gas composition and spectrum of the radiation field.

We performed our simulation using a 3D Cartesian domain of $(x,y,z) \in [-10.0\times 10^{18}, 10.0\times 10^{18}]$ cm, using five static mesh refinement levels. At the coarsest level, $n=0$, the entire domain $D$ is covered, and more refined levels with $n>0$ cover a sub-domain of diameter $D/2^n$, i.e. each finer level has a domain two times smaller than the level preceding it.
We included $256^3$ grid cells per level of refinement, such that the maximum grid resolution is $\Delta x = 4.88\times10^{15}$ cm. 

To simulate the stellar wind, we placed the RSG star at the origin, and the nested grids were centred on the star. Our aim is to study the ablation of the RSG wind, and thus a constant mass loss source term is applied, with a surface temperature $T_{\mathrm{eff}} = 3500~$K, a mass-loss rate \mdot\ = $2 \times 10^{-5}$ \Msun~ \yr$^{-1}$ \citep{Mackey2015} and wind terminal velocity $\varv_{\infty} = 25$ \kms \citep{Mauron2011}. The wind injection region is $5 \times 10^{16}$ cm, or ten cells. This is much larger than the stellar radius, and thus we injected the wind at the terminal velocity.

To simulate the cluster wind, we imposed a constant x-velocity  $V_x = - 1000$ km s$^{-1}$, density $\rho = 1\times10^{-24}~$ g cm$^{-3}$, and pressure $5\times 10^{-9}$ dyn cm$^{-2}$. While direct observational constraints are challenging, the cluster wind velocities cannot significantly exceed the adiabatic sound speed. This is of order 500-1000 km s$^{-1}$, using the diffuse X-ray gas temperatures measured by \cite{Kavanagh2011}. Our velocity is also consistent with that of \cite{2011ApJ...743..120S}, but lower than theoretical predictions  \citep[e.g.][]{Stevens2003,Haerer2023}, which are of order 2000-3000 km\,s$^{-1}$. Our density value is based on the density of the X-ray emitting gas from \cite{Kavanagh2011}. The upstream boundary condition was set to the fixed inflow conditions of the cluster wind, and the other external boundaries are set to zero-gradient (outflow).

The simulation was run for 200\,000 years with an explicit, finite-volume integration scheme that is 2nd-order accurate in time and space.
The radiative heating and cooling source terms are included by the operator-splitting method.
The timescale for the cluster wind to to advect from the RSG to the downstream boundary is $10^{11}$\,s (about 3000\,yr), much shorter than the simulation duration.
The advection timescale for the RSG wind to reach the downstream boundary is $4\times10^{12}$\,s (about 125\,000\,yr), and so we can expect the simulation to have reached a time-stationary state by 200\,000\,yr.

\section{Results}
\label{sec:results}

\subsection{XY slices}
\label{sec:results_slices}

\begin{figure}
    \centering
    \includegraphics[width=1\linewidth]{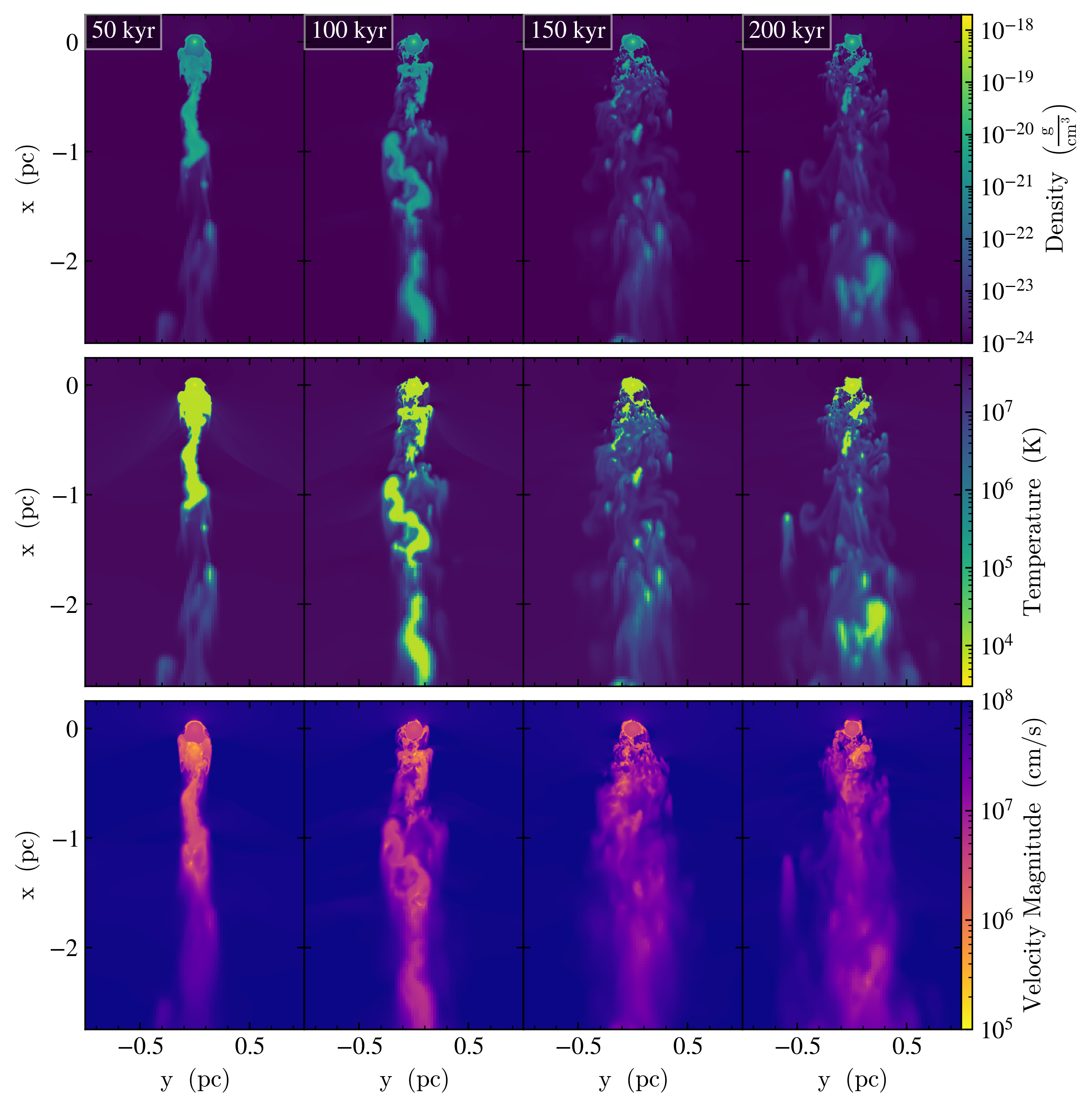}
    \caption{$x-y$ slices through our simulation of the plane $z=0$ at 50, 100, 150, and 200 kyr, showing distributions of density (upper), temperature (middle), and velocity magnitude (lower). The cluster wind is flowing from top to bottom, and the RSG is positioned at the origin. The colour scale is chosen to highlight dense, cool, and slow-moving material from the RSG.}
    \label{fig:end_xyxz_3panel}
\end{figure}
In Fig.\ref{fig:end_xyxz_3panel} we show $x-y$ slices of gas density, temperature and velocity magnitude at 50 kyr intervals. The cluster wind ablates the slowly expanding RSG wind. The contact discontinuity of the RSG wind and cluster wind becomes dynamically unstable in the upstream direction, with a combination of Rayleigh-Taylor (RT) and Kelvin-Helmholtz (KH) instability shearing clumps of shocked RSG wind into the wake downstream from the star. At early times, there is a dense and contiguous tail of ablated RSG material, probably influenced by initial conditions. This tail breaks up into clumps, and in the last two columns we have a clumpy and turbulent wake behind the star. These clumps are themselves being slowly ablated and their gas absorbed into the hot phase.

The ablation flow bears some similarity to the 2D von Karman vortex street, generated by subsonic flow past a solid object.
Direct estimates of Reynolds numbers (Re) in simulations of this type are difficult due to the requirement for artificial viscosity. Since we are in between a laminar flow (Re $\sim 10$) and fully turbulent (Re $\sim 1000$s), we can assume our Re is of order $\sim 100$s. With this assumption, for an idealised von Karman vortex street (where the characteristic length scale, $D,$ is the diameter of the cylinder), the shedding frequency, $f,$ is related to the Strouhal number (St) by the ratio of $D$ (in our case, twice the stand-off distance to the apex of the bow shock) and flow velocity $v$. 
For flow past a cylinder $v$ is well defined and, importantly, subsonic, but in our case we have the complication of two different flow speeds (the cluster wind and the RSG wind) with widely differing sound speeds.
The difficulty in assigning a characteristic velocity in such a multi-scale shear flow is addressed by \citet{Tan2021}.
If we nevertheless assume $v$ corresponds to the RSG wind velocity, we can calculate a characteristic timescale.
As St only varies slightly for Re $= 10^2-10^5$, we assumed St = 0.2 and find a typical shedding timescale of $\sim$ 3 kyr. However, due to the instability of the bow shock, $D$ is neither fixed nor constant over the bow shock, and so the vortices are only quasi-periodic. This is similar to the behaviour found by \cite{WarZijOBr07} in simulations of asymptotic giant branch stars, albeit with much lower relative velocities.

\subsection{Phase diagrams}
\label{sec:results_phase}

\begin{figure}
    \centering
    \includegraphics[width=1\linewidth]{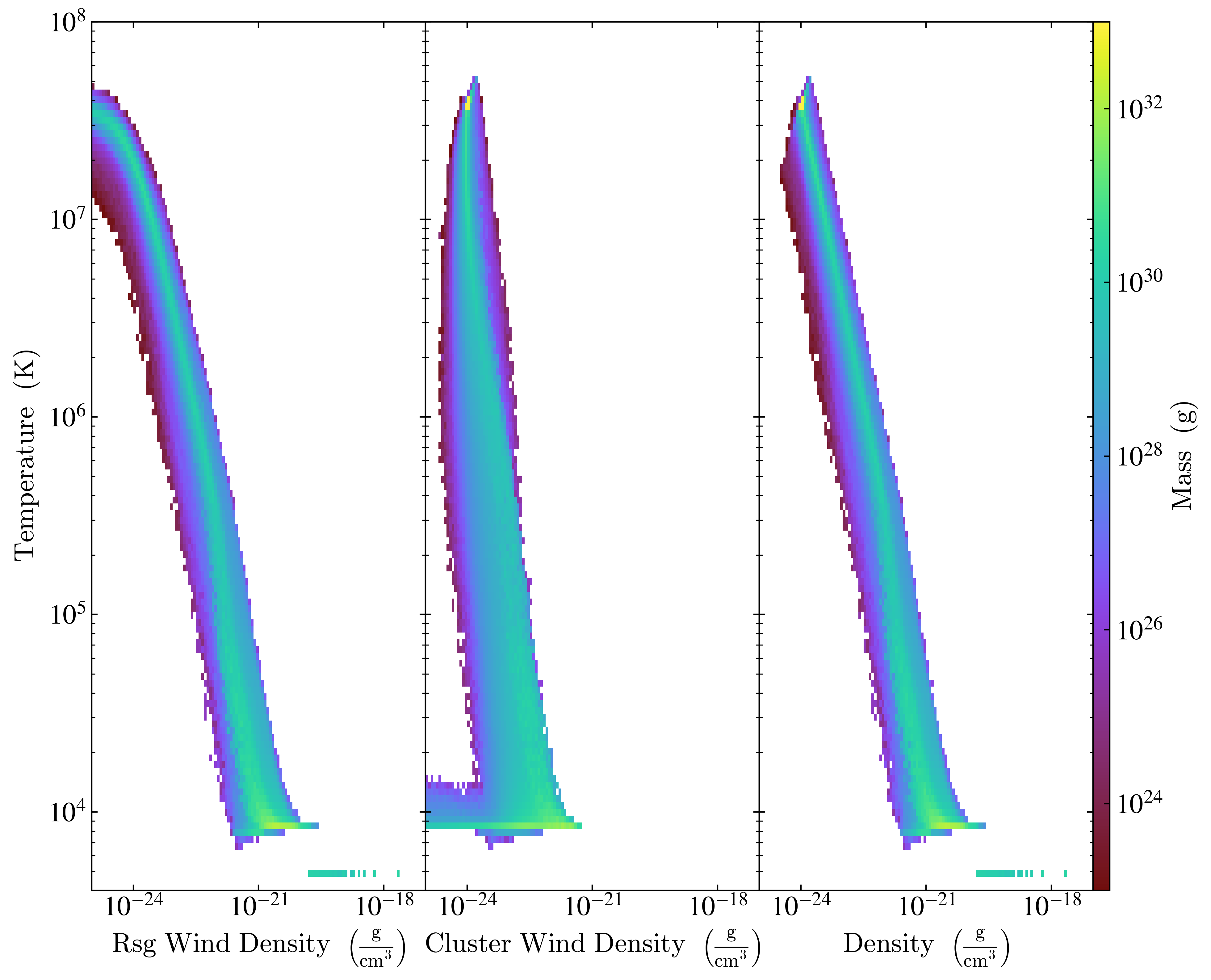}
    \caption{Temperature--density phase diagrams for the 3 pc sphere centred at the origin of our model for the RSG wind (left), cluster wind (middle), and all material (right) at the end of our simulation.}
    \label{fig:wind_ism_phase_end}
\end{figure}

In Fig. \ref{fig:wind_ism_phase_end} we show the density-temperature phase diagrams for only the RSG wind, only the cluster wind, and the sum of these components at the end of our simulation. We considered the RSG wind to be `cool', material above the photoionisation equilibrium temperature $\sim 8300$K to be `mixed', and material significantly hotter than this to be `hot'. In particular, RSG wind material that is over $10^4$K must have been heated by interactions, shocks and other processes.

Focusing on the RSG wind first, there is a horizontal feature at low temperature that corresponds to material immediately surrounding the star, mainly unshocked stellar wind. At the bow shock, the RSG wind is ablated by the cluster wind and clumps break off. As seen in Fig.\,\ref{fig:end_xyxz_3panel}, these clumps can remain dense and cool. This explains the `hotspot' around $\sim 10^{-20}$ g cm$^{-3}$ and the equilibrium temperature of $\sim 8300$K. As these clumps continue to be mixed, some of the material is heated and produces the long tail in that panel. 

For the cluster wind in the centre panel of Fig.\,\ref{fig:wind_ism_phase_end}, there is a  hotspot at the ambient cluster wind values of $\rho = 1\times10^{-24}~$ g cm$^{-3}$ and $T \sim 4\times 10^7$K. As the cluster wind mixes with the RSG wind, some of this material is cooled. Much of the mixed material cools until it reaches the equilibrium temperature ($\sim$$8300$K), and this is seen as a horizontal feature in the phase diagram. The presence of a feature at the equilibrium temperature in all three panels reflects the dynamical mixing of the two gas phases, which both heat the RSG wind and cool the cluster wind. 

\begin{figure}
    \centering
    \includegraphics[width=1\linewidth]{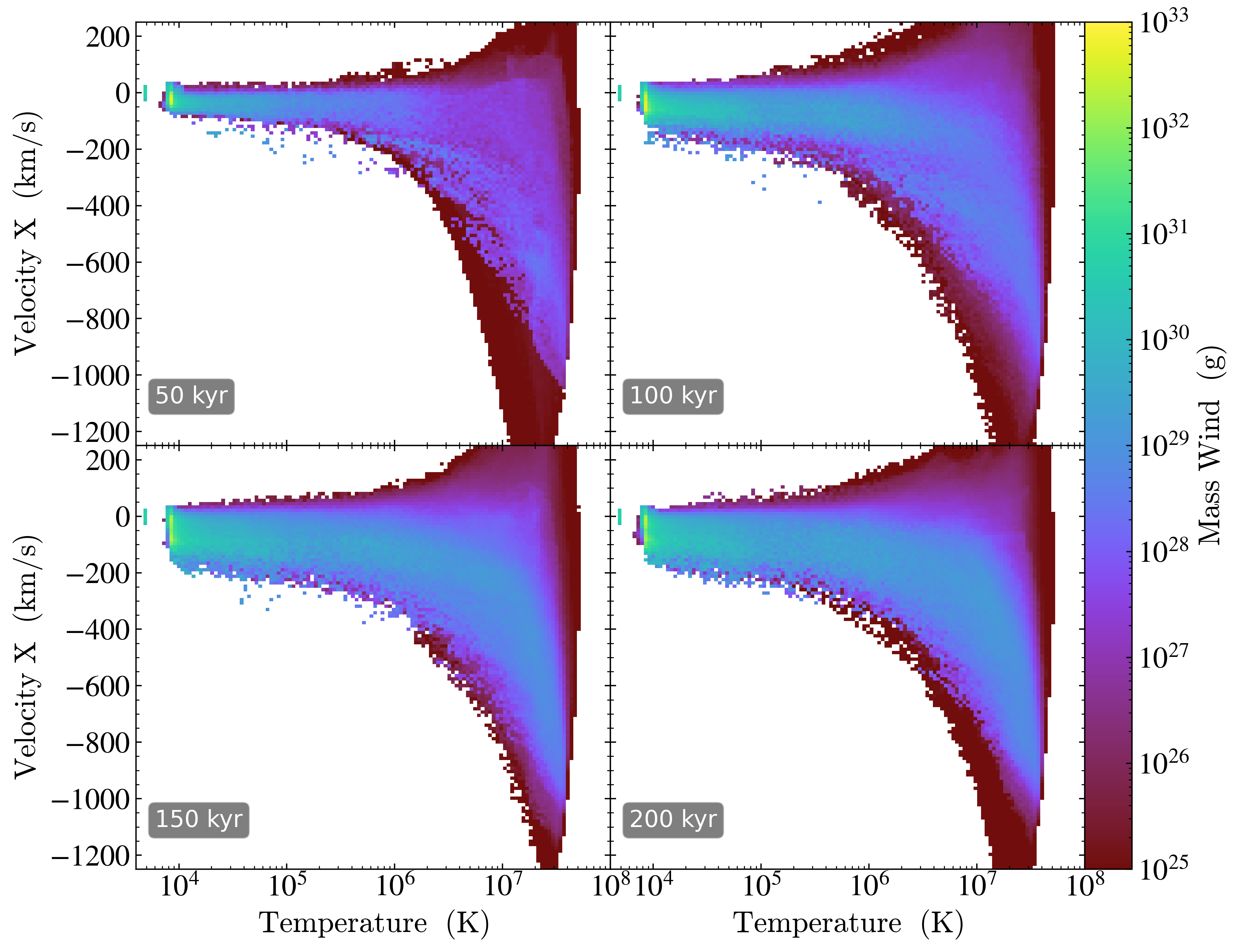}
    \caption{$V_x$--temperature phase diagrams for the 3 pc sphere centred at the origin of our model for only the RSG wind material for $t =$50, 100, 150, and 200 kyr.}
    \label{fig:timeseries_vx_t}
\end{figure}

In Fig.\,\ref{fig:timeseries_vx_t} we show phase diagrams of $V_x$ versus temperature for the RSG wind material at 50 kyr intervals. There is a hotspot at the wind injection temperature, visible as the isolated point in the upper left in all panels. As the RSG wind material expands spherically, we see a vertical feature where the material reaches the equilibrium temperature with a dispersion in velocity at the bow shock, which widens over time as the simulation approaches a time-stationary state.

The RSG wind material is ablated by the cluster wind, and rapidly accelerated as it is heated and then exits the domain, and the effect increases with time. The similarity of the last two panels implies that the simulation is approaching a time-stationary state.

\subsection{Mass loading}
\label{sec:results_ML}

\begin{figure}
    \centering
    \includegraphics[width=1\linewidth]{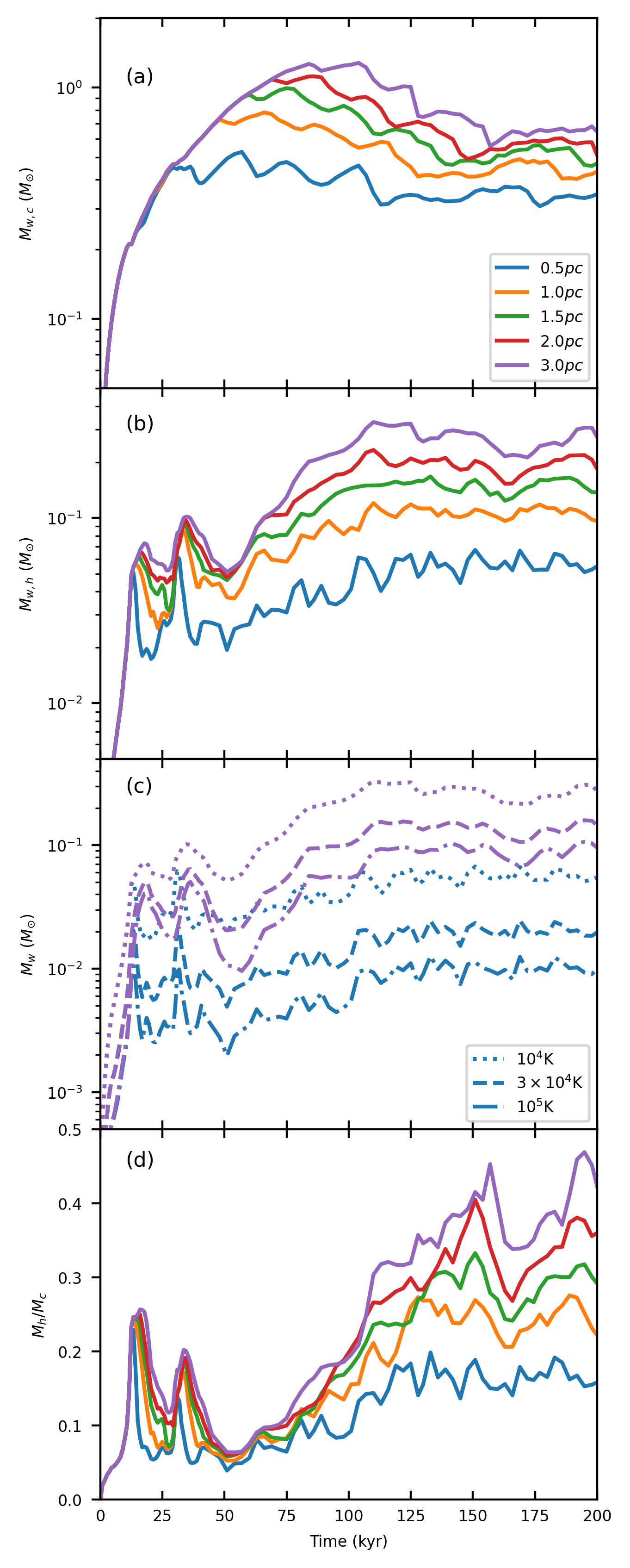}
    \caption{Time evolution of stellar wind masses of different temperatures enclosed within spheres of selected radii. Panel a: Cool ($T < 10^4$K) material. Panel b: Hot ($T > 10^4$K) material. Panel c: Material with $T > 10^4$K (dotted), $T > 3\times10^4$K (dashed), and $T > 10^5$K (dot-dashed) enclosed within spheres of $0.5\,$pc (blue) and $3.0\,$pc (purple). Panel (d): Ratio of cool ($T < 10^4$K) to hot ($T > 10^4$K) material. }
    \label{fig:hot_cold_mass_t}
\end{figure}

To get an idea of the efficiency of the mass loading over time, we compared the time evolution of  `hot' $M_\text{h}$ ($T > 10^4$K) and `cool' $M_\text{c}$ ($T < 10^4$K) RSG wind material (see Fig.\,\ref{fig:hot_cold_mass_t}). As expected, the cool RSG wind mass increases rapidly at early times. It stays relatively constant over time for small radii, but appears to peak and then decrease for larger radii (Fig.\,\ref{fig:hot_cold_mass_t}a). As the contiguous tail is broken up and ablated, the cool material is removed from the domain more rapidly over time. 
The mass of hot RSG wind material changes rapidly at first (Fig.\,\ref{fig:hot_cold_mass_t}b), likely due to initial conditions. At later times it tends to increase to a quasi steady state for all radii. As this material is accelerated and exits the domain, we only find a lower limit for the amount of RSG material that is heated. 
This behaviour is consistent across different heating temperatures at small and large radii (Fig.\,\ref{fig:hot_cold_mass_t}c). The small fluctuations arise from the turbulent nature of the instabilities themselves. 

The ratio of hot to cool RSG wind material tends to increase over time, again with turbulent fluctuations. Given that some of the heated wind is escaping from the domain, this represents a lower limit on the mass loading. Taking $M_\text{h}/M_\text{c} \approx 0.5$, this represents a mass-loading efficiency of 33\% at late times, albeit a more detailed numerical study with a larger simulation domain is required to measure this number accurately. We study the numerical convergence of these results in Appendix \ref{app:res_test}.

Second generation star formation from rapidly cooling shocked stellar winds has been studied in detail by \cite{2017ApJ...835...60W}. However, that requires material to be shielded (i.e. not be photoionised), which is never the case in the central part of Westerlund 1 that we are considering due to the magnitude of the incident ionising radiation. We therefore do not expect there to be second generation star formation resulting from the ablation flows in this case.

\begin{figure}
    \centering
    \includegraphics[width=1\linewidth]{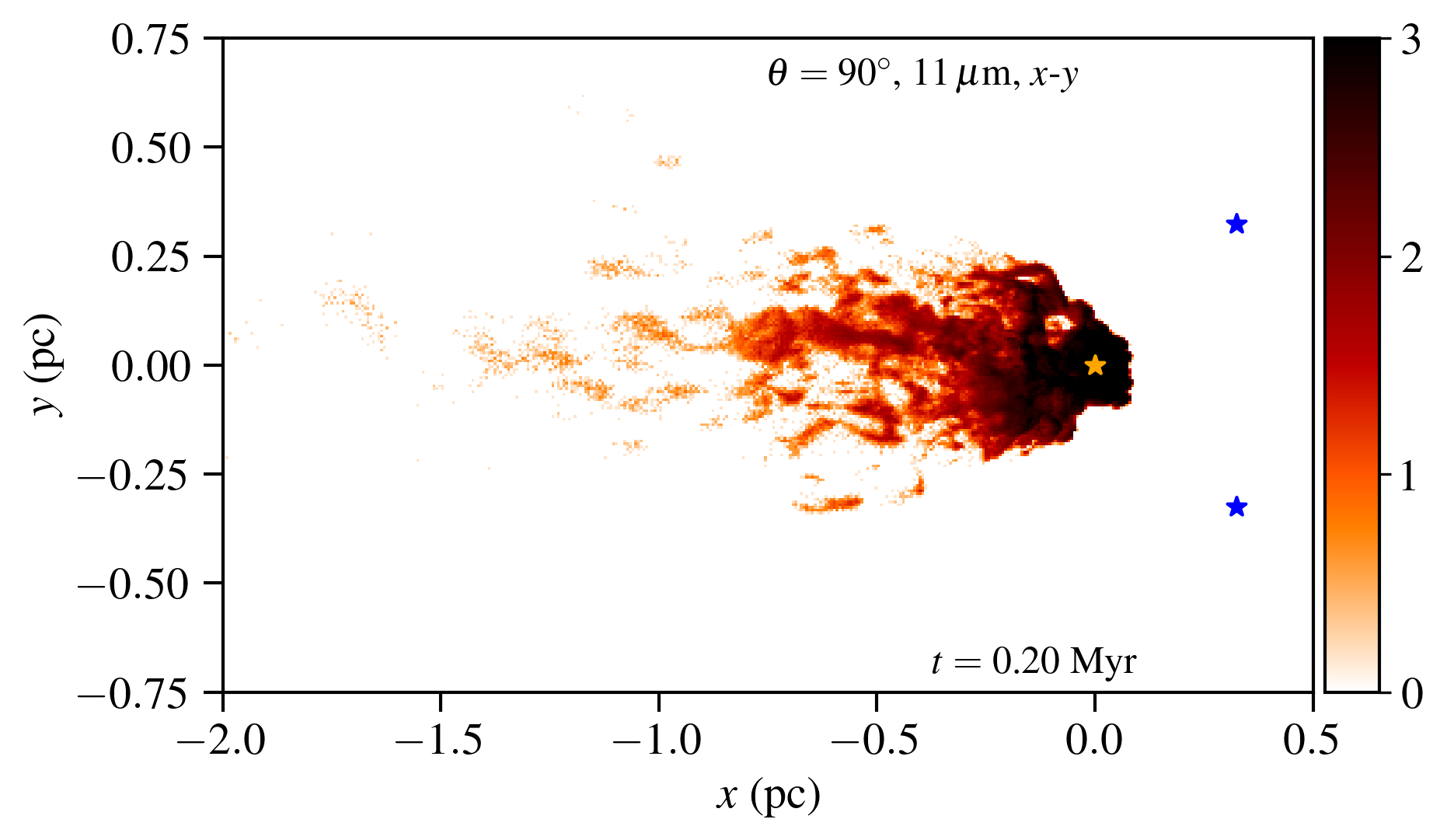}
    \includegraphics[width=1\linewidth]{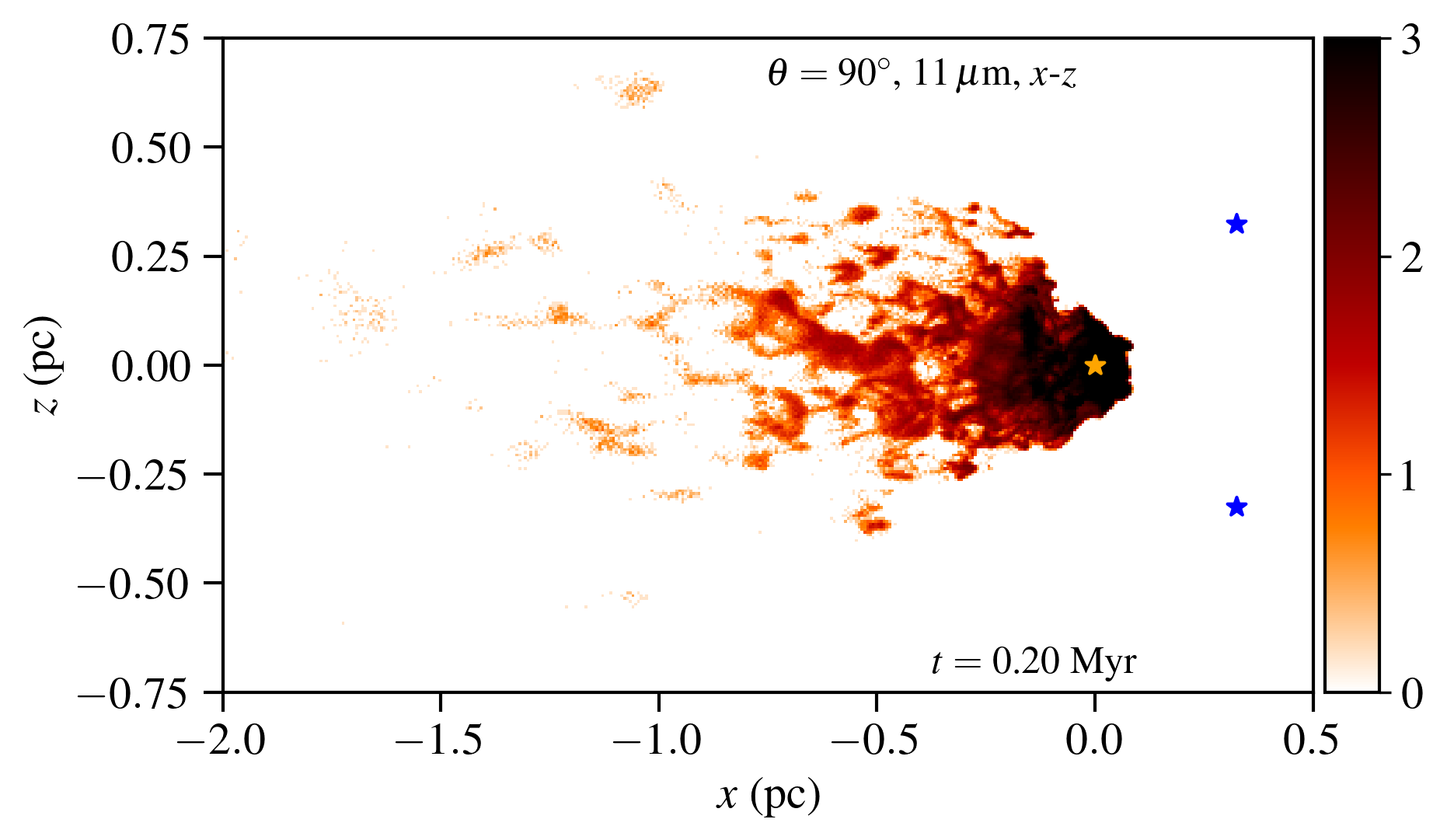}
    \caption{Thermal dust emission maps from the final snapshot of our simulation, showing surface brightness at 11\,$\mu$m on a logarithmic scale.
    The colour bar shows $\log_{10} I_{11\,\mu\mathrm{m}} / (\mathrm{MJy\,ster}^{-1})$.
    The radiation sources heating the dust are the RSG at the origin (orange star) and two sources indicated by the blue stars at $x=0.32$\,pc and $z=\pm0.32$\,pc.}
    \label{fig:11um-maps}
\end{figure}

\begin{figure}
    \centering
    \includegraphics[width=1\linewidth]{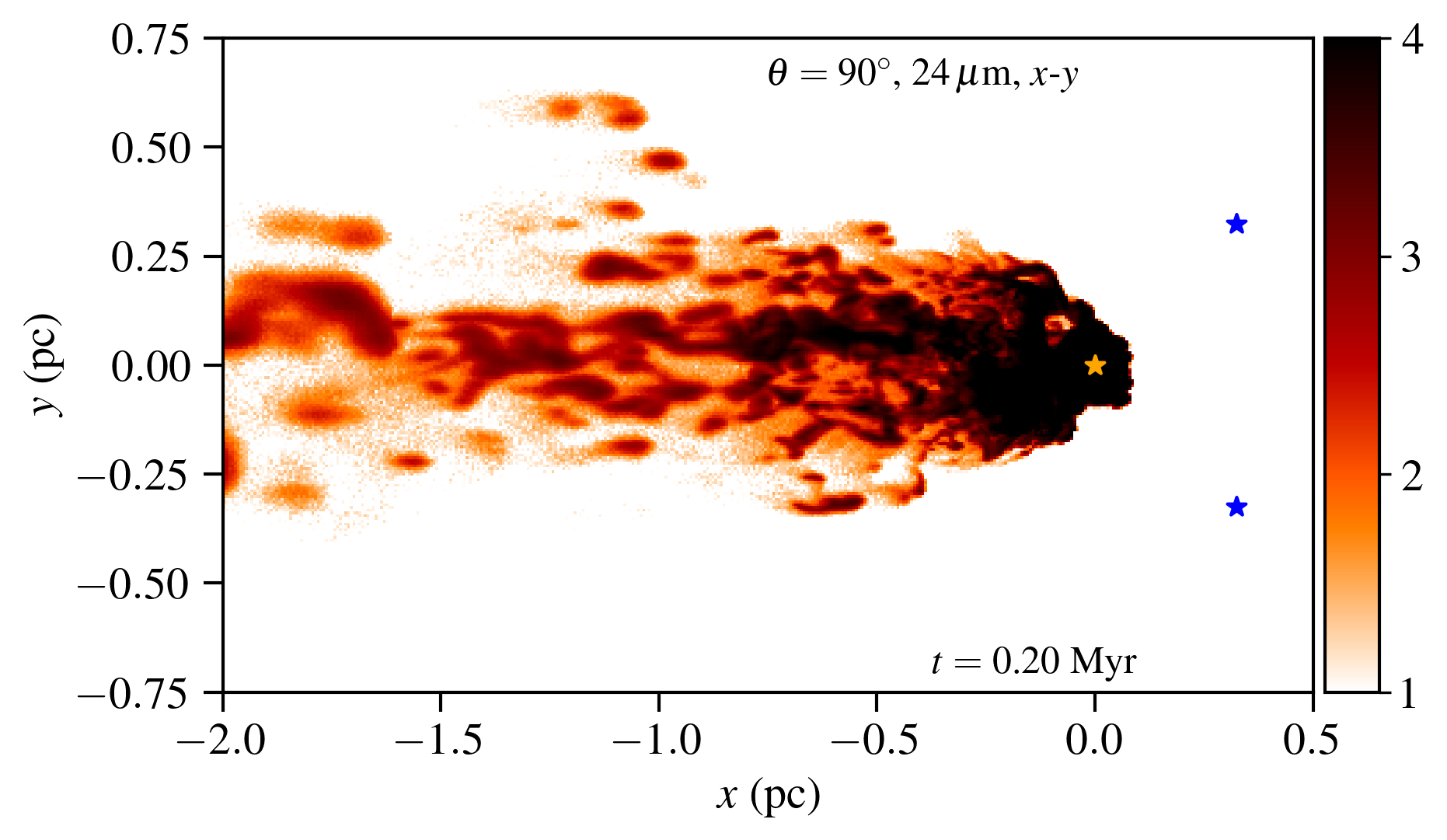}
    \includegraphics[width=1\linewidth]{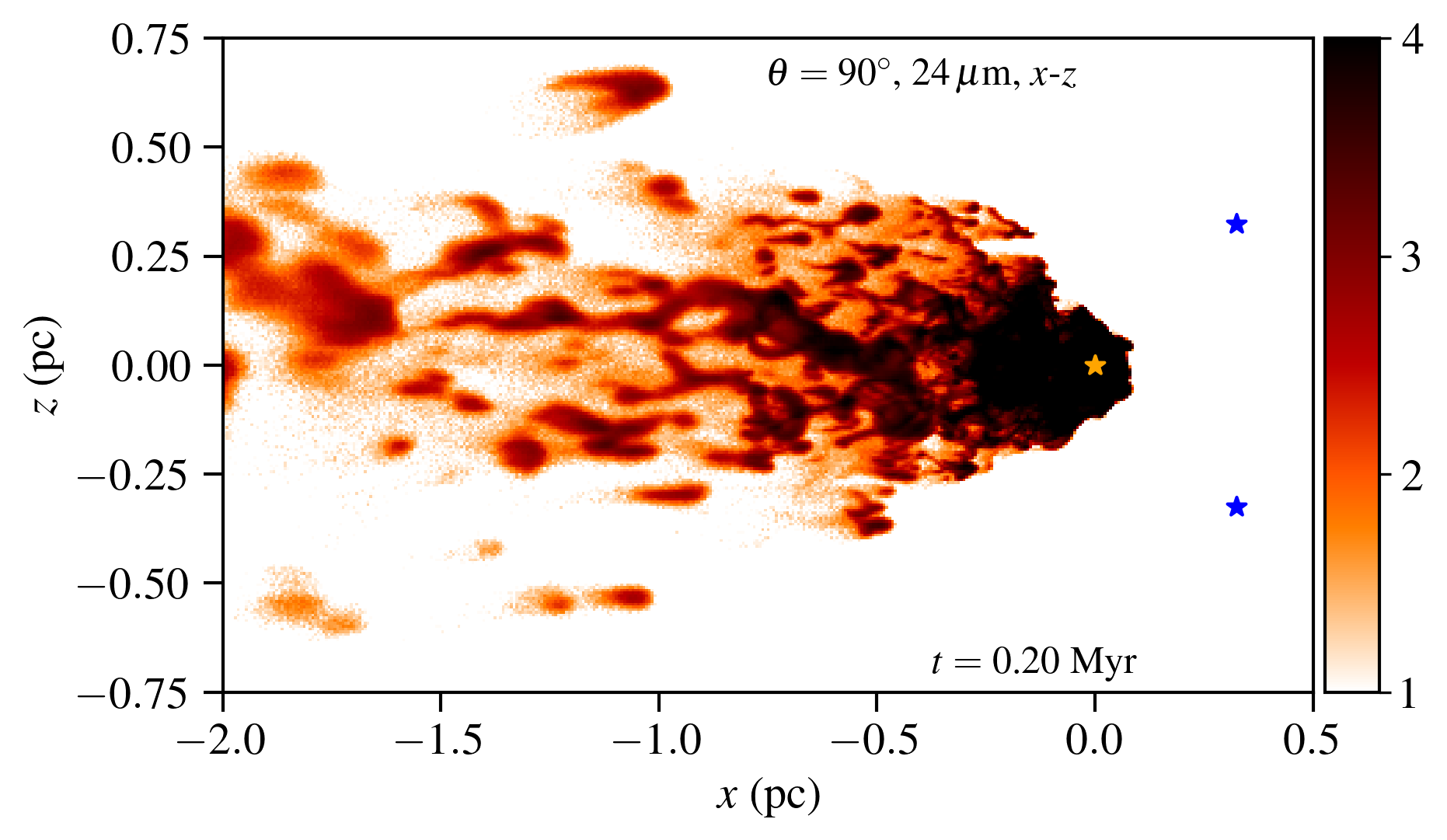}
    \caption{Same as Fig.~\ref{fig:11um-maps} but showing surface brightness at 24\,$\mu$m on a logarithmic scale.
    The colour bar shows $\log_{10} I_{24\,\mu\mathrm{m}} / (\mathrm{MJy\,ster}^{-1})$.}
    \label{fig:24um-maps}
\end{figure}

\begin{figure*}
    \centering
    \includegraphics[width=\textwidth]{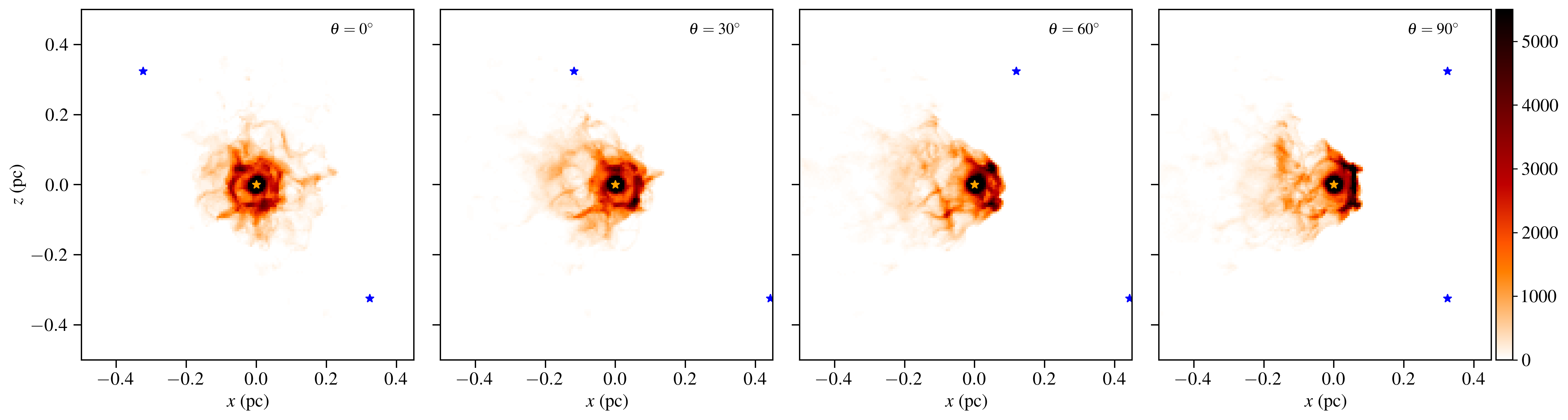}
    \includegraphics[width=\textwidth]{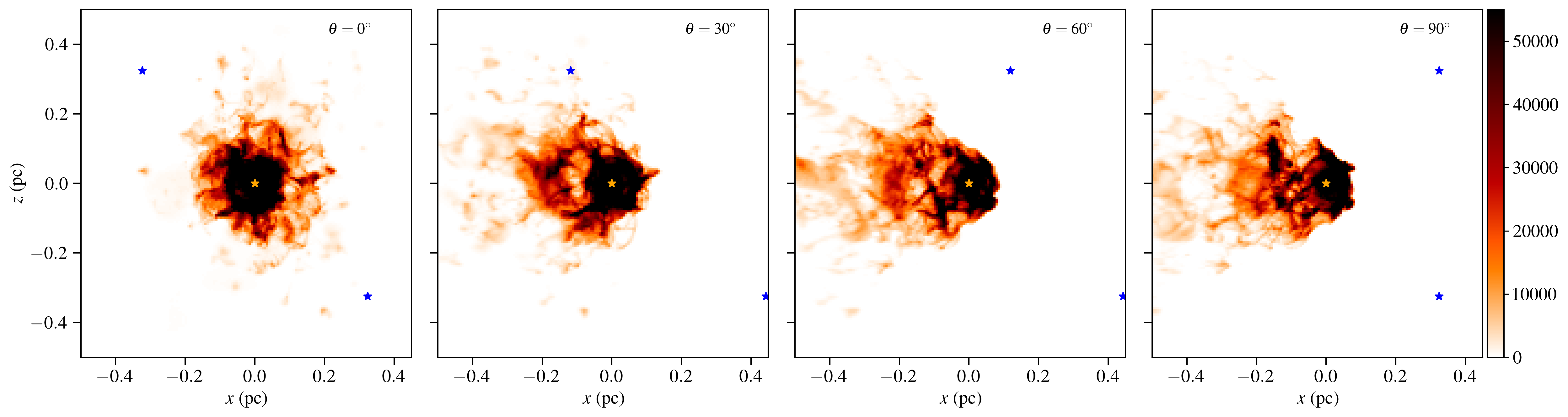}
    \caption{Synthetic thermal dust emission maps at 11\,$\mu$m (top row) and 24\,$\mu$m (bottom row) viewed from different angles, with emission shown on a linear scale in units of $\mathrm{MJy\,ster}^{-1}$. 
    The location of the RSG is shown by the orange star, and the other two synthetic radiation sources by the blue stars projected at $z=\pm0.32$\,pc.
    The angle $\theta$ is measured between the observer's line of sight and the cluster-wind flow velocity.
    }
    \label{fig:11um-angles}
\end{figure*}

\subsection{Dust maps}
\label{sec:results_dust}

We produced synthetic dust maps using the \textsc{TORUS} Monte Carlo radiative transfer code \citep{Harries2019} with the procedure outlined in \citet{2021PION} to post-process our simulation.
To account for the external radiation field that contributes to dust heating, we included the RSG itself as one source, and placed two hot-star sources at positions $[1,1,1]\times10^{18}$\,cm and $[1,-1,-1]\times10^{18}$\,cm.

The spectral energy distribution (SED) of the two hot-star sources are assumed to be blackbodies with an effective temperature $T_\mathrm{eff}=40$\,kK, radius $20\,\mathrm{R}_\odot$, and a luminosity $L_\star=9.1 \times 10^5 \, \mathrm{L}_\odot$.
While this is more luminous than estimated values for individual stars in Westerlund 1 \citep[the supergiants are at least a factor of three less luminous;][]{ClaNegCro05, BeaDavSmi21}, the motivation for these parameters is to generate a radiation field comparable to that found in the inner regions of the Westerlund 1 cluster. In this way we avoid the complications of including tens of radiation sources with different positions and photospheric properties, making the simplest possible assumptions for this first study of the subject.
At the location of the RSG, the resulting radiation energy flux from the two sources is 92.4\,erg\,cm$^{-2}$\,s$^{-1}$ and radiation energy density is $1.9\times10^3$\,eV\,cm$^{-3}$. Further downstream the radiation flux decreases with the inverse square law, so the radiative heating is reduced.

For the RSG we took a \citet{Kurucz1992} stellar atmosphere model with $T_\mathrm{eff}=3750$\,K and stellar radius $1000\,\mathrm{R}_\odot$ with a stellar mass of $25\,\mathrm{M}_\odot$. This corresponds to a luminosity of $1.78\times10^5\,\mathrm{L}_\odot$. The RSG radiation field heats the dust in the immediate vicinity of the star, but in the downstream region the heating is dominated by the two hot radiation sources, particularly because the dust absorption cross-section peaks in the UV.
We assumed silicate grains \citep{Dra03} with a dust-to-gas ratio of 0.01, and a \citet{MatRumNor77} grain-size distribution with power-law index of -3.3 from 0.005 to 0.2\,$\mu$m. We further assumed that grains do not exist in gas with temperatures $T>10^6$\,K, i.e. instantaneous grain destruction. This is a gross simplification of the rich diversity of dust observed in winds of RSGs \citep[e.g.][]{Dec21}, but serves to obtain a reasonable estimate of the dust temperature and emissivity in the RSG wind. It would be relatively easy to obtain intensity variations of a factor of two or more by changing the dust composition, size distribution and dust-to-gas ratio within reasonable values, and so the predicted intensity maps have relatively large modelling uncertainties.

We present synthetic dust continuum-emission maps for 11\,$\mu$m in Fig.~\ref{fig:11um-maps} and 24\,$\mu$m in Fig.~\ref{fig:24um-maps}, as these are of interest to compare to current and possible future observations of RSG outflows in Westerlund 1 \citep{Guarcello2024}. The upper panel in each figure shows a projection with the line of sight being the $z$-axis, and the lower panel shows the same but along the $y$-axis. The logarithmic intensity scale brings out some of the faint features further in the downstream wake of the flow. The faintest features are starting to show some of the photon-sampling noise that is inherent in Monte Carlo radiative transfer modelling.

The brightest emission at both wavelengths is concentrated around the RSG, where the radiation field is most intense and the dust has the highest temperature.
Because the peak of the dust SED is at $\lambda>11\,\mu$m, and the $11\,\mu$m band is in the Wien tail of the SED, the emission in Fig.~\ref{fig:11um-maps} is about $10\times$ fainter than at 24\,$\mu$m (Fig.~\ref{fig:24um-maps}).
Moving to the downstream wake (more negative $x$ coordinates) the dust temperature decreases further with distance from the radiation sources, and so there is very little $11\,\mu$m emission for $x<-0.75$\,pc.
The 24\,$\mu$m emission is less sensitive to distance but eventually the dust temperature decreases such that this band is also in the Wien tail of the SED and so the brightness decreases significantly.

We see a range of clumps, globules with and without tails, and some filamentary emission connecting clumps that are being ablated from the RSG wind.
These features are somewhat reminiscent of models of bow shocks from winds of cool runaway stars \citep{WarZijOBr07, MohMacLan12}, although here the much larger flow velocity of the cluster wind (and the higher resolution of our simulations) leads to a more highly clumped wake behind the star.
The half opening-angle of the IR-emitting clumps in the downstream is about $30^\circ-40^\circ$, with the majority of the dusty wind material enclosed within a cylinder of radius $\approx0.5$\,pc.
It should be noted, however, that we imposed a planar inflow rather than a spherically expanding cluster wind (which is relevant on scales of a few parsecs), and so we probably underestimate the dispersion of the clumps in directions perpendicular to the cluster-wind flow ($-\hat{x}$).
The optical depth of the clumps to EUV/far-UV and optical stellar radiation is $\tau\approx0.075-0.3$ (depending on clump size and photon energy), calculated using the opacity tables from \textsc{torus}, and so they are not self-shielding and are photoionised throughout to a high degree.

Projections of the 11\,$\mu$m and 24\,$\mu$m emission from different viewing perspectives are shown in Fig.~\ref{fig:11um-angles}, progressing from face-on ($\theta=0^\circ$, where $\theta$ is the angle between the line of sight and the cluster-wind flow velocity) to edge-on ($\theta=90^\circ$) from left to right.
The face-on images have a circular symmetry with a `fried-egg' morphology that results from a peak of emission in the termination-shock of the RSG wind.
As $\theta$ increases, the nebula acquires the head-tail morphology of an ablation flow, with dynamical instabilities and a turbulent wake.
At 11\,$\mu$m, Fig.~\ref{fig:11um-angles} shows an emission feature that resembles a bow shock from the edge-on perspective, although the cluster wind is only trans-sonic and no emission from it is visible in dust maps. This feature is the shocked RSG wind, which accumulates in a dense isothermal layer behind the RSG wind termination-shock and then advects downstream around the unshocked RSG wind region.

\subsection{Comparison to observations}
\label{sec:results_comp}
We compared our synthetic observations with the JWST/MIRI F1130W observations towards Westerlund 1 by \cite{Guarcello2024}, taken as part of GO-1905 (PI Guarcello). Since our objective is a phenomenological comparison rather than bespoke modelling of a specific system, we used the level 3 data products from the MAST\footnote{\url{https://archive.stsci.edu/}} science archive without further processing or reduction. 

\begin{figure}
    \centering
    \includegraphics[width=1\linewidth]{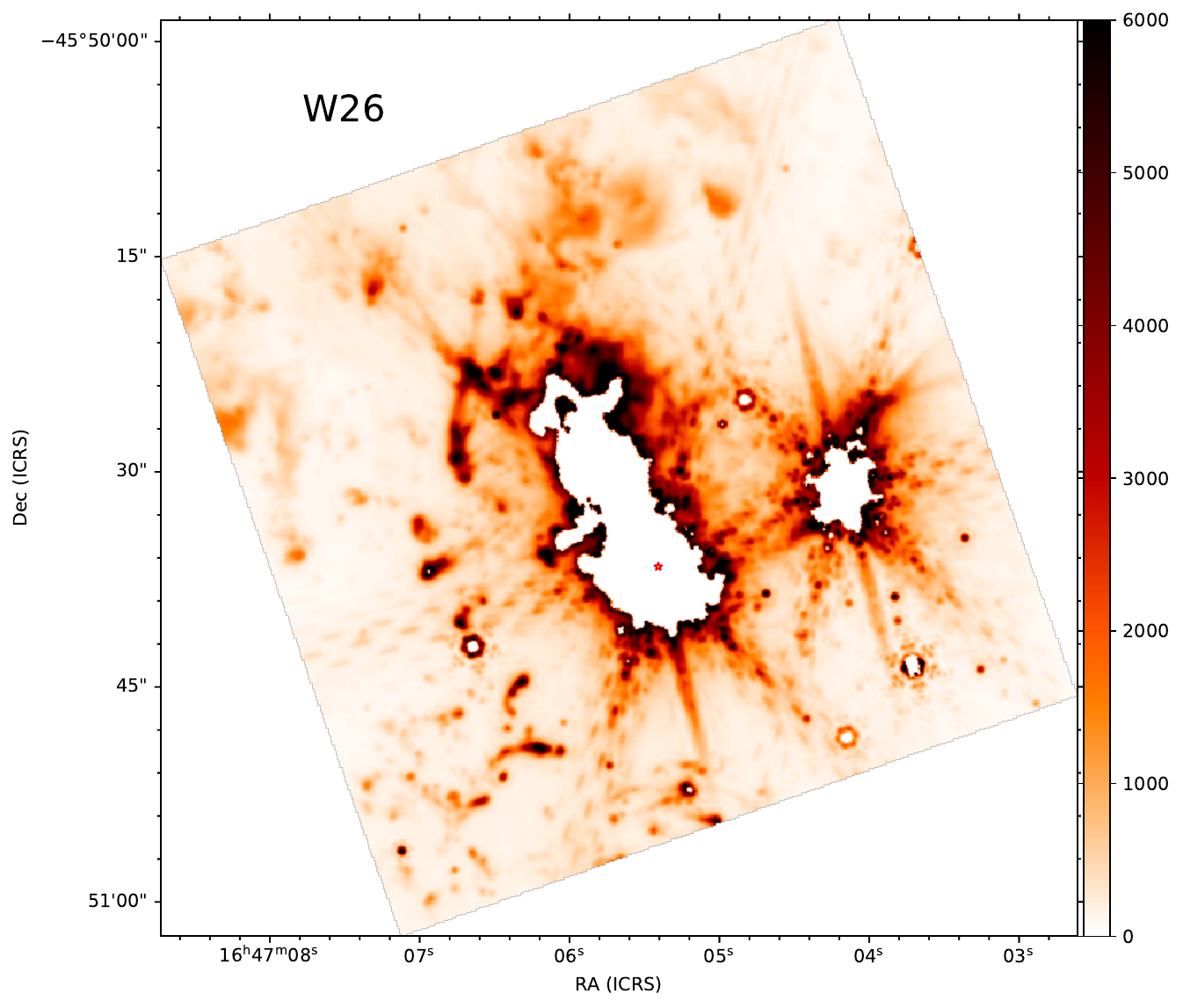}    
    \includegraphics[width=1\linewidth]{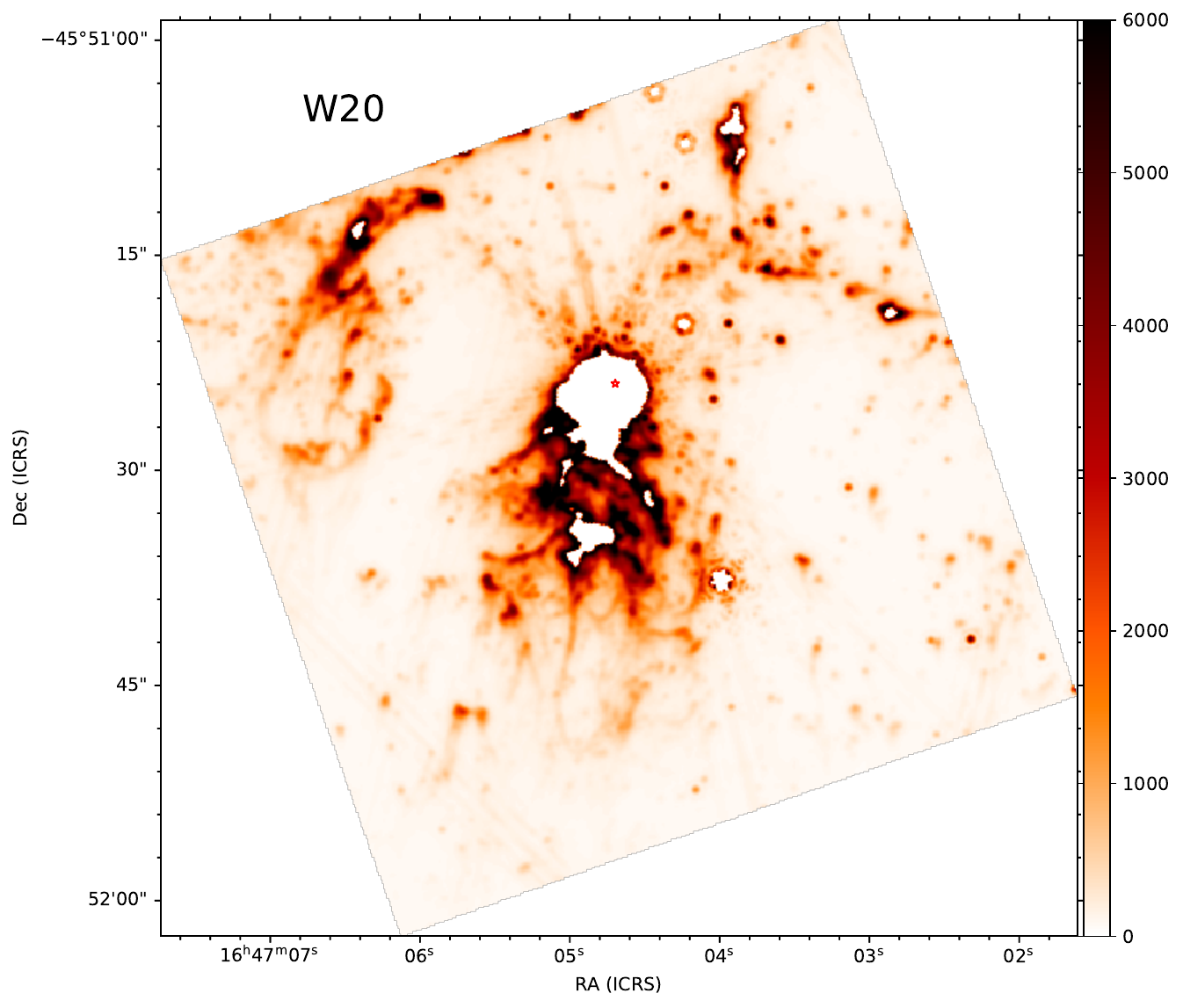}
    \includegraphics[width=1\linewidth]{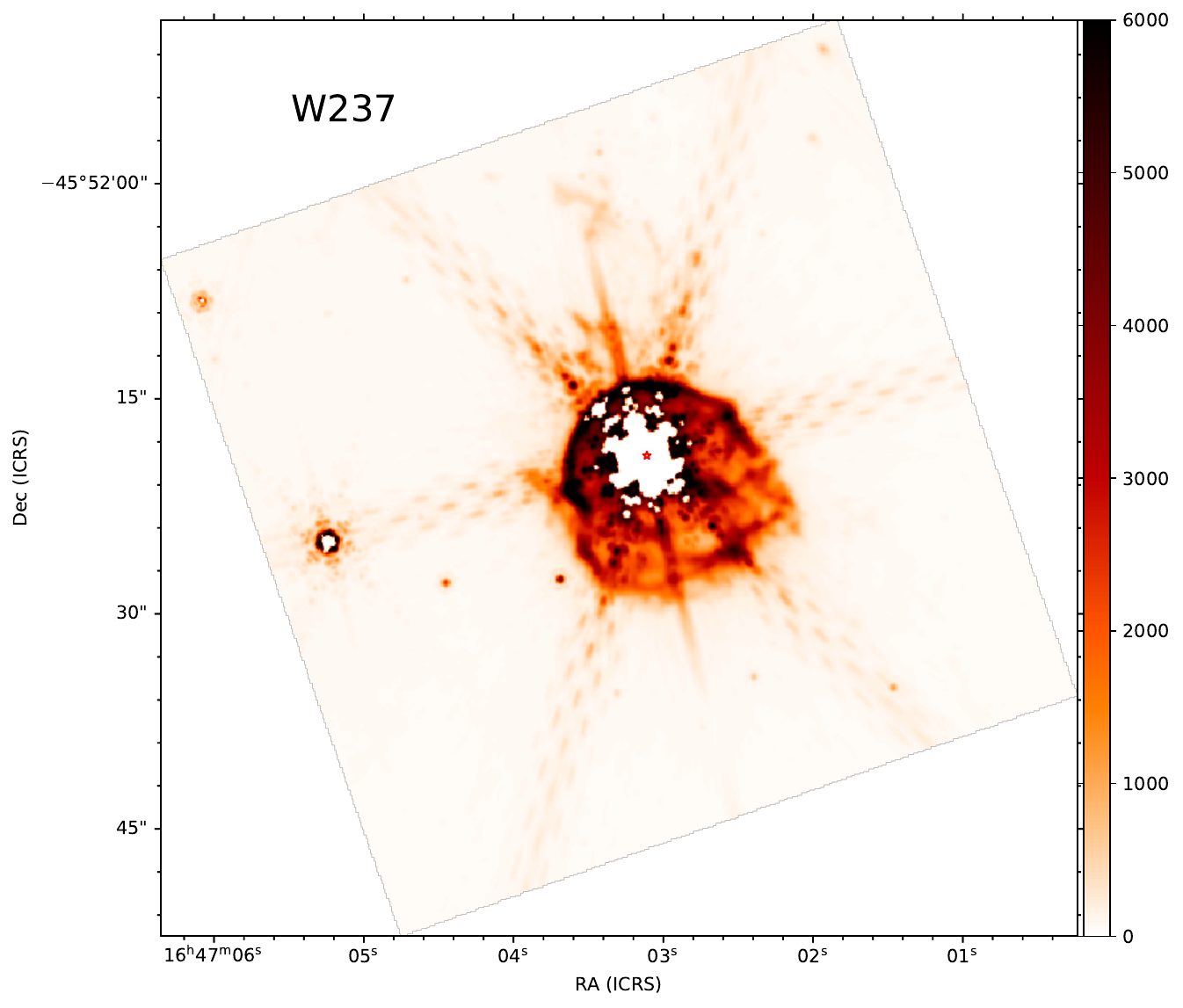}    
    \caption{JWST/MIRI F1130W observations towards three M-type supergiants in Westerlund 1. These are W26, W20, and W237 from top to bottom and in order of decreasing declination. The colour bar units are MJy\,ster$^{-1}$. In each case the central white pixels are saturated. The stars denote the positions of the three target supergiants. The colour scale is the same as the synthetic observations in Fig. \ref{fig:11um-angles}.}
    \label{fig:F1130W}
\end{figure}

We focused on three M-type supergiants: W20, W26, and W237, which are plotted in Fig.\ref{fig:F1130W}. 
In each case there is a substantial saturated inner zone, simply due to the high sensitivity of JWST and bright nature of the sources. However, exterior to that is a plethora of unsaturated features.

Qualitatively and quantitatively, the 11\,$\mu$m synthetic image from Fig.~\ref{fig:11um-angles} at $\theta=90^\circ$ is very similar to the JWST F1130W image of the RSG W237 (\citealt{Guarcello2024}; see also our Fig.~\ref{fig:F1130W}).
The synthetic images at $\theta=60^\circ$ and $90^\circ$ are also remarkably similar to the F1130W image of the RSG W20, although there the JWST image seems to show brighter emission that our prediction.
For both W20 and W26, a detailed comparison with simulations is difficult because large parts of the nebula are saturated in the F1130W image.

The synthetic 24\,$\mu$m emission maps in Fig.~\ref{fig:11um-angles} show emission that is more than ten times brighter than at 11\,$\mu$m, as discussed already for Fig.~\ref{fig:24um-maps}.
Plotting with a linear rather than logarithmic intensity scale brings out the chaotic and filamentary morphology of the wake behind the star, with clumps connected by ridges of dusty gas as the ablation process rips them away from the bulk of the RSG wind material.

\begin{figure*}
    \centering
    \includegraphics[width=0.9\linewidth]{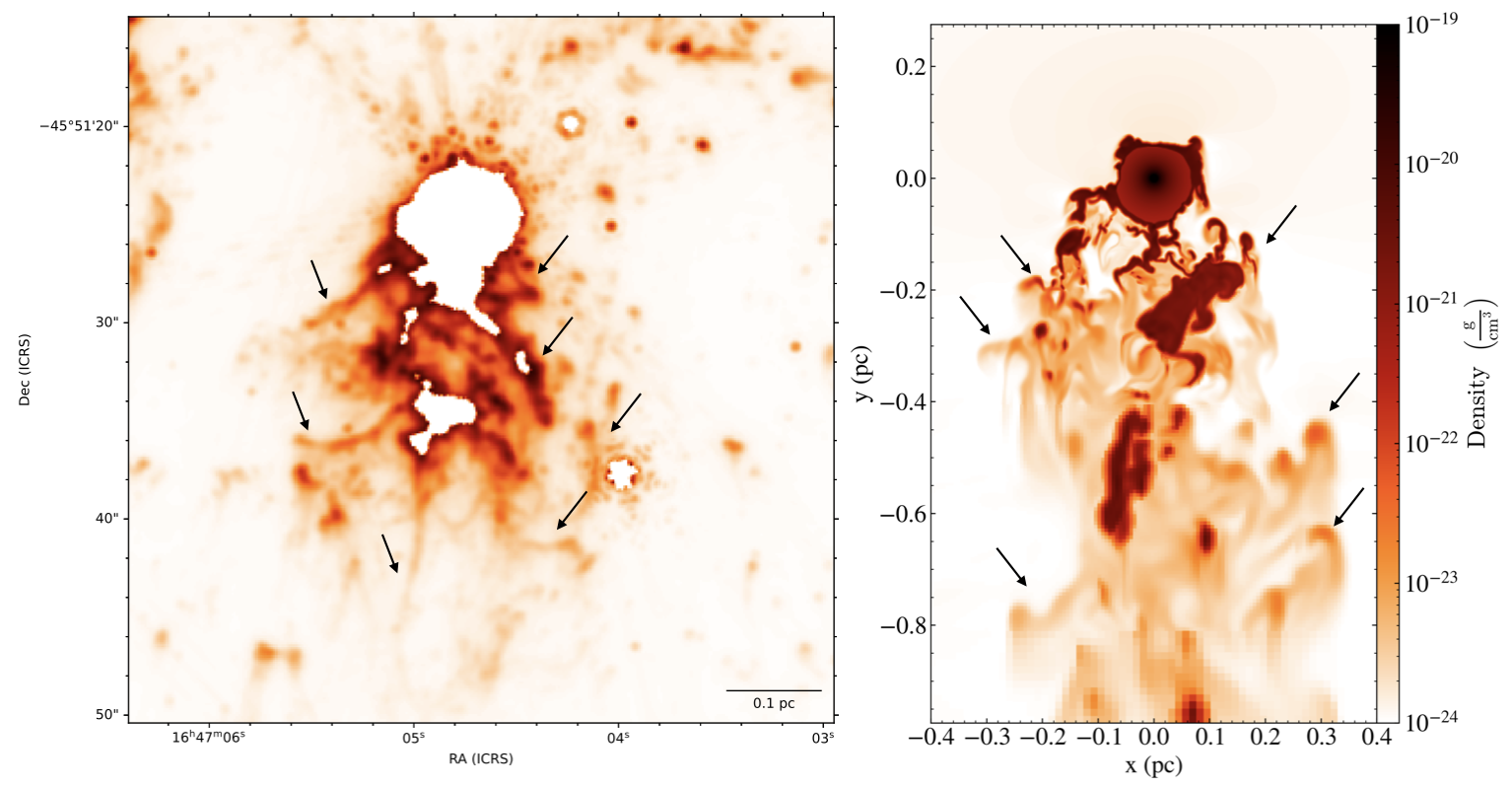}
    \caption{Left: JWST/MIRI F1130W observations towards W20. Right: Simulation density map. In addition to the primary flow and cometary knots, there are extended features, indicated by the arrows, that connect to the primary flow and arise due to RT and/or KH instabilities.  }
    \label{fig:morphologycomparison}
\end{figure*}

The fluxes in the unsaturated parts of the extended features associated with the M-type supergiants are typically in the range 3000-8000\,MJy\,ster$^{-1}$, comparable to (i.e. same order of magnitude as) the values in the simulated maps of Figs. \ref{fig:11um-maps}, \ref{fig:24um-maps}, and \ref{fig:11um-angles}. The morphology of the flow is also similar, with a primary flow, multiple cometary secondary objects and even extended features that connect to the primary flow that likely result from some combination of RT and/or KH instabilities, as illustrated in Fig.\ref{fig:morphologycomparison}. 
Such features have previously been discussed in the context of runaway asymptotic giant branch stars such as Mira \citep{WarZijOBr07}, although here the numerical resolution is significantly higher and a more complex flow structure is apparent.

It is notable that the model predicts similar fluxes to the observations without any contribution from polycyclic aromatic hydrocarbon (PAH) emission, despite there being a significant PAH C-H bending feature within this band \citep[e.g.][]{2006ApJ...653..267T, 2015MNRAS.448.2960C, 2024MNRAS.528.3392W}. This suggests that the PAH abundance in the ablation flow is depleted, most likely due to destruction by shocks or radiation \citep[e.g.][]{2010A&A...510A..36M, 2010A&A...511A...6S}. We note that PAHs are not universally present in RSG winds \citep{Verhoelst2009} and so it is possible that they are not formed in the first place. Future observations may be able to resolve this question.

This appears to be supported by ALMA continuum observations of free-free emission from the nebulae around the cool supergiants in Westerlund 1 \citep{2018A&A...617A.137F}, ATCA radio observations that also detect extended nebulae around the cool supergiants \citep{2018MNRAS.477L..55A}, and H$\alpha$ emission from the nebula around W26 \citep{2014MNRAS.437L...1W}, all implying that the winds of the cool supergiants are externally photoionised. However, confirming a deficit of PAH abundance will require a multi-wavelength study with dedicated models that treat PAHs, which is beyond the scope of this work. 

\section{Discussion}
\label{sec:discussion}

\subsection{RSG and cluster wind uncertainties}
\label{sec:discussion_uncert_wind}
The mass-loss rates of RSGs are poorly constrained \citep{2020MNRAS.492.5994B} and can vary rapidly over time, including strongly clumped mass-loss episodes \citep{2021Natur.594..365M}. In this work we chose a constant `canonical' value, but a time-varying mass-loss rate would increase the variation of the mass-loading rate and shedding timescale, as well as the morphology of dust emission. 
The cluster wind properties are also poorly constrained from observations. To fully account for this, a model of the cluster wind's velocity, density, and temperature as a function of radius would be required. We also assumed a planar inflow here, whereas in reality it is spherically expanding, with potentially strong local perturbations from nearby individual stars. The cluster wind expansion radius is of the order of the cluster radius itself, and therefore could impact the cluster wind at the positions of the RSGs. 
We see the effects of our initial conditions for the first 25-50\,kyr of our simulation, as we `switch on' our star. In reality, the RSG would have evolved from a main sequence star with a hot, fast wind. The RSG would expand into the bubble excavated during the main sequence as opposed to the uniform ISM we assumed here. 

\subsection{Modelling uncertainties}
\label{sec:discussion_uncert_model}

Many previous works have focused on the physics of radiative turbulent mixing layers, which arise in many different astrophysical contexts \cite[e.g.][]{Fielding2020,Tan2021,Mackey2025}. \cite{Fielding2020} show that while the phase structure is resolution dependent, the total cooling is well converged even for moderate resolutions. We did not consider the effects of magnetic fields in this work. We would expect the morphology of the dust filaments to be elongated along the cluster magnetic field lines \citep{Gronke2020} but the contribution to the turbulent mixing is likely not significant \citep{Li2020}. We did not consider thermal conduction in this work either. The importance of thermal conduction is debated in the literature, but recent work suggests that its effects are only significant if larger than the turbulent diffusivity \citep{Ji2019,Tan2021}. 

Our dust modelling has several uncertainties. We assumed a generic hot radiation field, but the dust emission is sensitive to the specifics of this radiation field. For example, a hot star in the negative-$x$ direction would heat up the dust and change the observed dust map considerably. We also made assumptions about the properties of the dust grains, and the ratio of gas to dust in the ablation flow. Given the number of uncertainties and the fact that we did not tune any parameters, the qualitative and quantitative agreement between our synthetic 11$\,\mu$m emission maps and the observed emission around the RSGs in Westerlund 1 is encouraging. 

\subsection{Mass loading}
\label{sec:discussion_ML}

We obtain a lower limit for mass-loading efficiency due to RSG wind material being accelerated and exiting the domain before it can be heated, as well as due to resolution as we discuss in Appendix \ref{app:res_test}. Using the cluster wind velocity formalism given in Eq.\,1 of \cite{Haerer2023}, and assuming a total current RSG mass-loss rate in Westerlund 1 of $1\times 10^{-4}$ \Msun~\yr$^{-1}$ , we would expect the cluster wind to be slowed by a factor of $\sim$$10\%$ due to RSG winds. 
Westerlund 1 is a young cluster with a small number of RSGs at present, so here this contribution is probably not important. At later times however, when a cluster wind is weaker due to fewer WR stars and mass loading is stronger due to more RSG stars, this effect could become more important. 
Furthermore, the JWST observations \citep{Guarcello2024} show that there is significant mid-IR nebular emission in all directions around the star cluster, some of which cannot have come from the RSGs unless the cluster-wind geometry is very unusual.  This material also contributes to the mass loading of the cluster wind, and so the combined effects of all of the cool, dusty gas is larger than the 10\% effect we estimate for just the RSG winds.

In this work we only considered a single set of steady-state stellar wind parameters, and future work is required to examine the effects of initial stellar mass and stellar evolution on mass loading in clusters. We also neglected other possibilities such as bursts and asymmetries in mass-loss, which would affect the morphology of the ablation structures and dust emission.

\subsection{Relevance to super star clusters}
\label{sec:discussion_SSC}
Super star-clusters are extremely massive YMCs that are sufficiently massive and tightly bound that they could be progenitors of globular clusters, which are often observed to have multiple stellar populations \citep{Gratton2012, 2018ARA&A..56...83B}.
The enrichment patterns of the second-generation stars imply that they formed from gas that was polluted by nuclear-processed gas ejected from the first generation of (massive) stars.
Feedback from (hot) massive stars is one explanation for these multiple populations \cite[e.g.][]{Wuensch2008,deMink2009,Krause2013}, and cool supergiants have also been proposed \citep[e.g.][]{2018A&A...612A..55S}.
Rather specific conditions are required in order for winds of hot stars to cool efficiently and remain bound to the star cluster to form a second generation \citep{2017ApJ...835...60W}.
RSG winds are already cool, dense and dusty, and thus readily suited to forming a second generation of stars if the wind material can be confined within the cluster.  While Westerlund 1 may not be massive enough to retain its gas and form a 2nd generation in situ, it is very interesting to study (presumably) cool and dusty gas right in the core of the cluster.
Measuring the thermal and chemical properties of this gas, and estimating its lifetime in the cluster, may give insights into how the multiple populations in globular clusters could have formed.
This is especially interesting because some targeted searches for intra-cluster gas and dust failed to detect either in significant quantities in other star clusters \citep{2014MNRAS.443.3594B}.

\subsection{Follow-up observations}
\label{sec:discussion_obs}

Given the high velocities associated with the ablation flow, of order 1000\,km\,s$^{-1}$, they should be spectrally resolvable even at medium resolution. VLT/ERIS-SPIFFER \citep{2023A&A...674A.207D} is a near-IR integral field spectrograph facility with a high resolution K-middle filter including Br\,$\gamma$ \citep[e.g.][]{2024MNRAS.527.3220R}.
This should therefore be well suited to confirming those high velocities without the same susceptibility to saturation that JWST suffers. That would immediately rule out a primordial origin for the nebular emission seen towards Westerlund 1 and confirm that it instead results from stellar ejecta. It would also provide the ability to make empirical mass-loss rate estimates, which is a key factor in interpreting these ablation flows. 

\citet{2010MNRAS.402..152W,2013MNRAS.435...30W} obtained spectra of dynamical mixing layers at the boundary between hot cluster winds and pillars of dense gas and dust. They were able to measure the enhanced velocity dispersion in the mixing layer of about 100\,km\,s$^{-1}$.
Similar observations of these dusty clumps in Westerlund 1 should give observational constraints on the degree of mixing between hot and cool phases, and hence on the mass loading of the cluster wind.
Comparison of such observations with synthetic spectra from HD simulations will be valuable for constraining physical processes operating in the mixing layer (e.g. HD mixing vs~thermal conduction).

\section{Conclusion}
\label{sec:conclusion}

In this work we computed a 3D HD model of a RSG embedded in a hot cluster wind with values typical for those in Westerlund 1. We summarise our findings below:

\begin{itemize}
    \item A clumpy tail of ablated RSG wind material forms through dynamical instabilities at the contact discontinuity, with the cluster wind shearing and entraining cold gas into the cluster wind. 

    \item The RSG wind material is heated as the clumps are further ablated in this tail, while some of the cluster wind is cooled due to dynamical mixing.

    \item The RSG wind material is accelerated as it is heated, and this acceleration tends towards a steady state over time.

    \item After any effects of the initial conditions have left the simulation domain, the fraction of cool RSG wind mass decreases over time while the fraction of hot RSG wind approaches a steady state. 

    \item We determine a mass-loading efficiency of about 33\% for the RSG wind, although this should be confirmed with higher-resolution simulations and for different RSG-wind and cluster-wind parameters.

    \item We produced synthetic dust continuum-emission maps for 11\,$\mu$m and 24\,$\mu$m for several viewing angles for comparison with current and future observations of RSG outflows in Westerlund 1 and other clusters.

    \item Our dust maps closely match observations of W20, W26, and W237 without any fine-tuning, in terms of both observed fluxes and morphology. Our omission of PAH emission in our dust map modelling suggests that PAHs are depleted in the ablation flow due to the effects of winds and radiation and/or a lack of PAH formation in the RSG winds in the first place.

    \item We encourage further observational studies of the fate of RSG winds and dusty clumps in clusters. This will help ascertain the role RSG winds play in mass loading, as well as their potential to seed future populations of stars, as inferred from observations of globular clusters.
    
    \item Follow-up observations with IR integral field spectroscopy such as with VLT/ERIS-SPIFFIER would enable detailed kinematic studies of the RSG--cluster wind interaction and resulting ablation flow. This would provide insights into the origin of Westerlund 1's nebular emission, a direct probe of the so-far elusive RSG mass-loss rates, and help disentangle the various physical processes in the subsequent mixing.
\end{itemize}

\begin{acknowledgements}
We thank the anonymous referee for their constructive feedback which has improved the article. The authors thank Brian Reville for useful discussions on hydrodynamics. The simulations presented here were performed on the HPC system Raven at the Max Planck Computing and Data Facility. CJKL gratefully acknowledges support from the International Max Planck Research School for Astronomy and Cosmic Physics at the University of Heidelberg in the form of an IMPRS PhD fellowship.
This publication results from research conducted with the financial support of Taighde \'Eireann - Research Ireland under Grant number 20/RS-URF-R/3712 (JM).
JM is grateful to B.~Reville and the Max Planck Institute for Nuclear Physics (MPIK) for welcoming him as a regular visitor during the preparation of this paper.
AACS is supported by the Deutsche Forschungsgemeinschaft (DFG, German Research Foundation) in the form of an Emmy Noether Research Group – Project-ID 445674056 (SA4064/1-1, PI Sander).
TJH acknowledges funding from a Royal Society Dorothy Hodgkin Fellowship and UKRI guaranteed funding for a Horizon Europe ERC consolidator grant (EP/Y024710/1). This research has made use of the Astrophysics Data System, funded by NASA under Cooperative Agreement 80NSSC21M00561. This study made use of the following software packages: 
     Astropy \citep{astropy:2018},  Numpy \citep{HarMilVan20}, matplotlib \citep{Hun07}, yt \citep{TurSmiOis11}, \textsc{pion} \citep{2021PION}, \textsc{pypion} \citep{GreMac21}, \textsc{TORUS} \citep{Harries2019}. \\

\textit{Contributions.} CJKL and JM conceived the project. CJKL performed and analysed the simulations and led the writing of the article. JM computed the \textsc{TORUS} dust maps, assisted with performing the simulations and interpreting the results. TJH created the JWST figures, assisted with \textsc{TORUS} and interpreting the results. AACS provided feedback on the manuscript and theoretical support for the article.   
\end{acknowledgements}

\bibliographystyle{aa} 
\bibliography{biblio.bib} 

\appendix
\section{Resolution test}\label{app:res_test}
\begin{figure}
    \centering
    \includegraphics[width=1\linewidth]{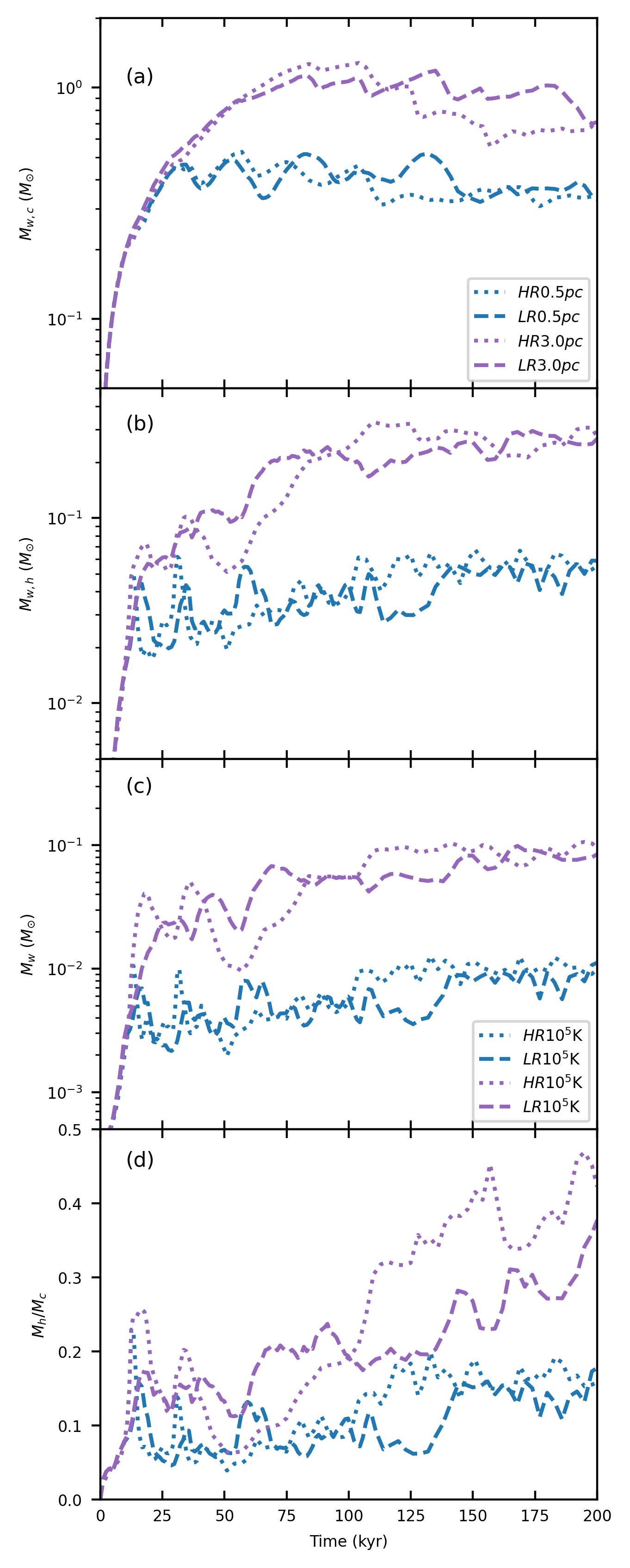}
    \caption{Time evolution of stellar wind masses of different temperatures for the low (dashed) and high (dotted) resolution models: Panel a: Mass of cool (T < $10^4$K) material vs time enclosed within spheres of 0.5 pc (blue) and 3.0 pc (purple). Panel b: Idem for hot (T > $10^4$K) material. Panel c: Plot of material with T > $10^4$K (blue) and T > $10^5$K (purple) enclosed within spheres of 0.5 pc (blue) and 3.0 pc (purple). Panel d: Ratio of cool (T < $10^4$K)  to hot (T > $10^4$K) material.}
    \label{fig:rescomp_hot_cold_mass_t}
\end{figure}

To check the effects of resolution on our results, we ran an otherwise identical model using 192$^3$ grid cells per level of refinement. PION has a minimum injection region radius of 7 cells, making this the lowest achievable resolution without having an artificially large injection region. We compare the time evolution of the RSG wind mass for the two models in Fig.\ref{fig:rescomp_hot_cold_mass_t}.

\begin{figure}
    \centering
    \includegraphics[width=1\linewidth]{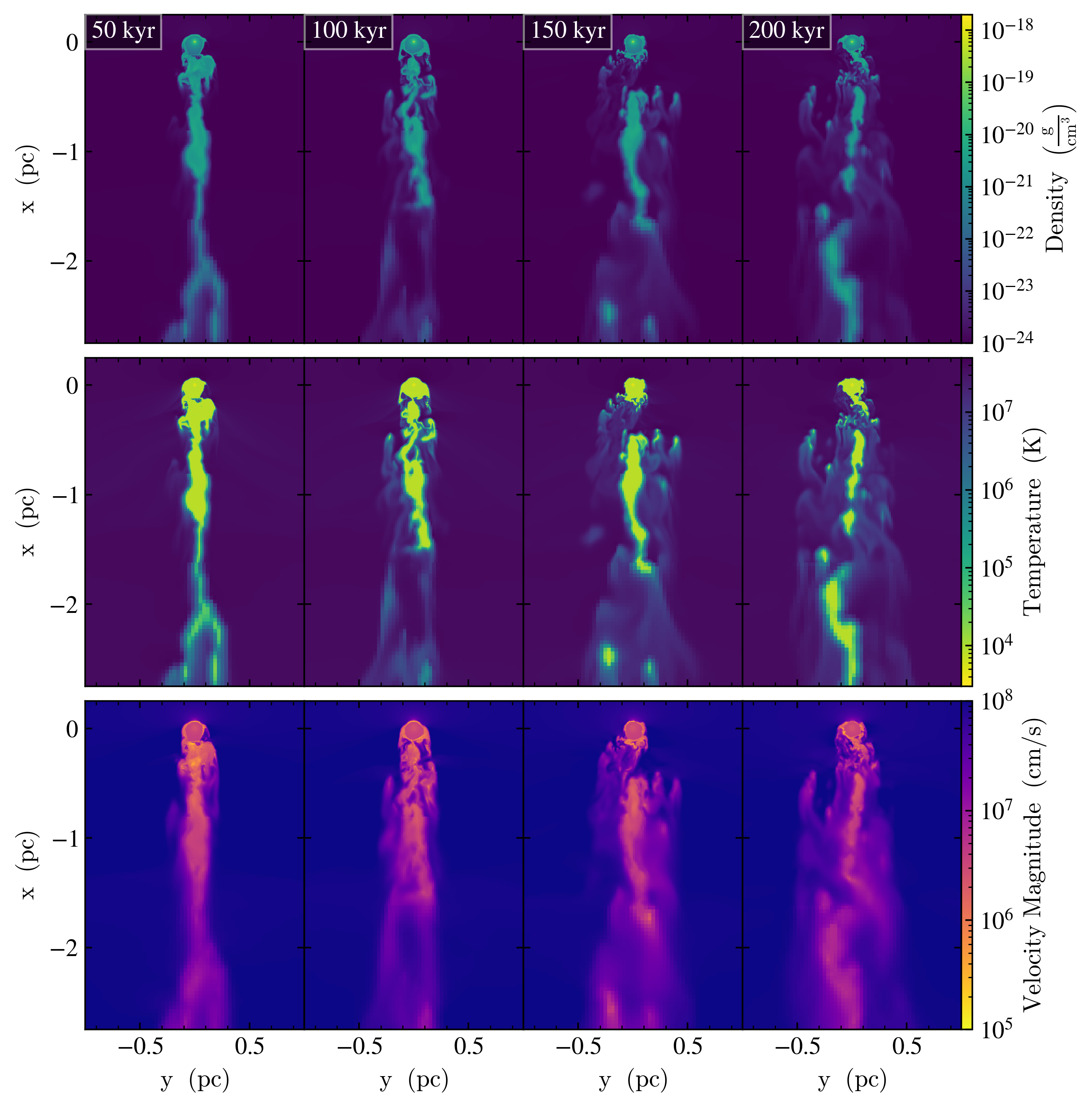}
    \caption{Same as Fig. \ref{fig:end_xyxz_3panel} but for the lower-resolution model.}
    \label{fig:app_end_xyxz_3panel}
\end{figure}

\begin{figure}
    \centering
    \includegraphics[width=1\linewidth]{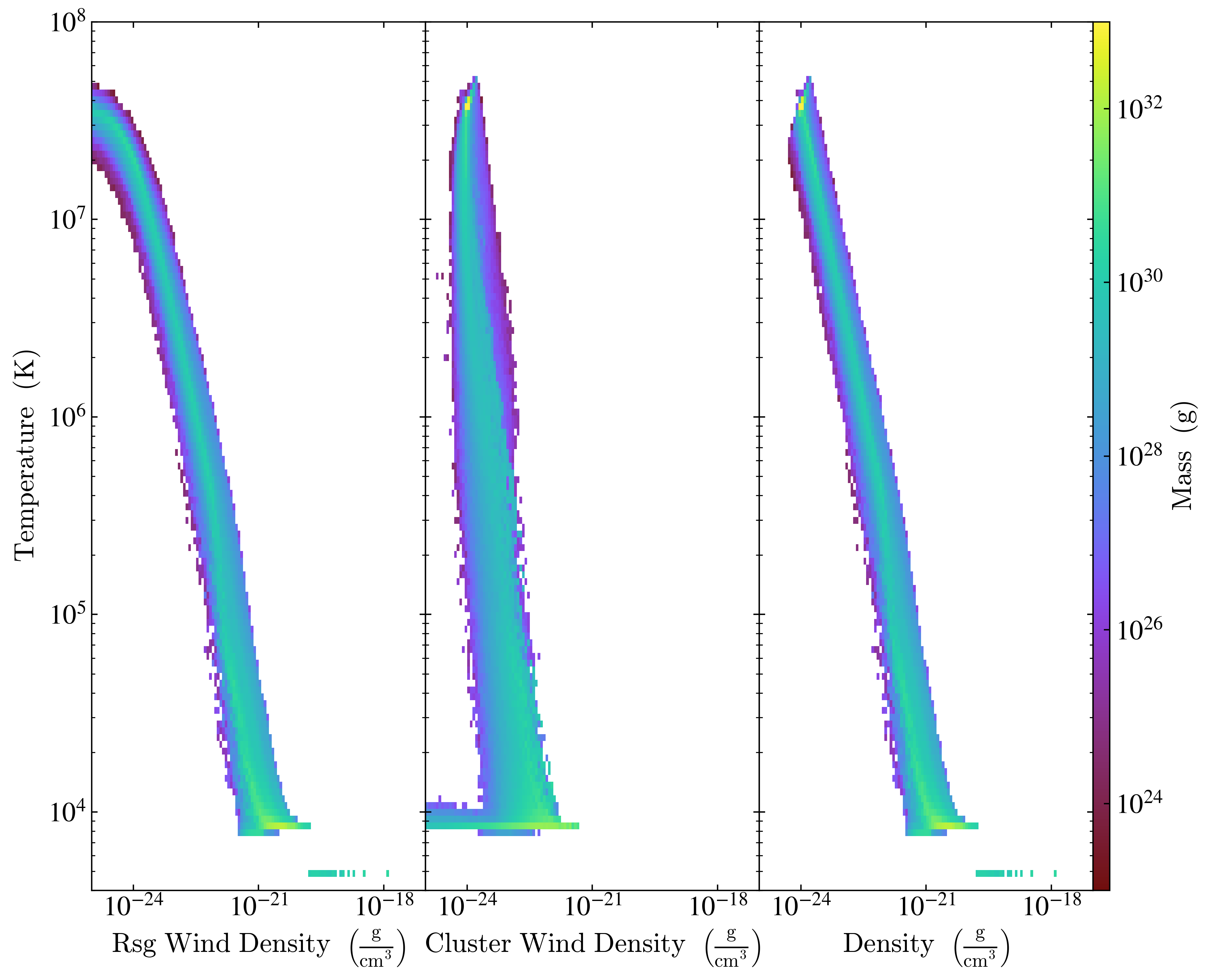}
    \caption{Same as Fig. \ref{fig:wind_ism_phase_end} but for the lower-resolution model.}
    \label{fig:app_wind_ism_phase_end}
\end{figure}

\begin{figure}
    \centering
    \includegraphics[width=1\linewidth]{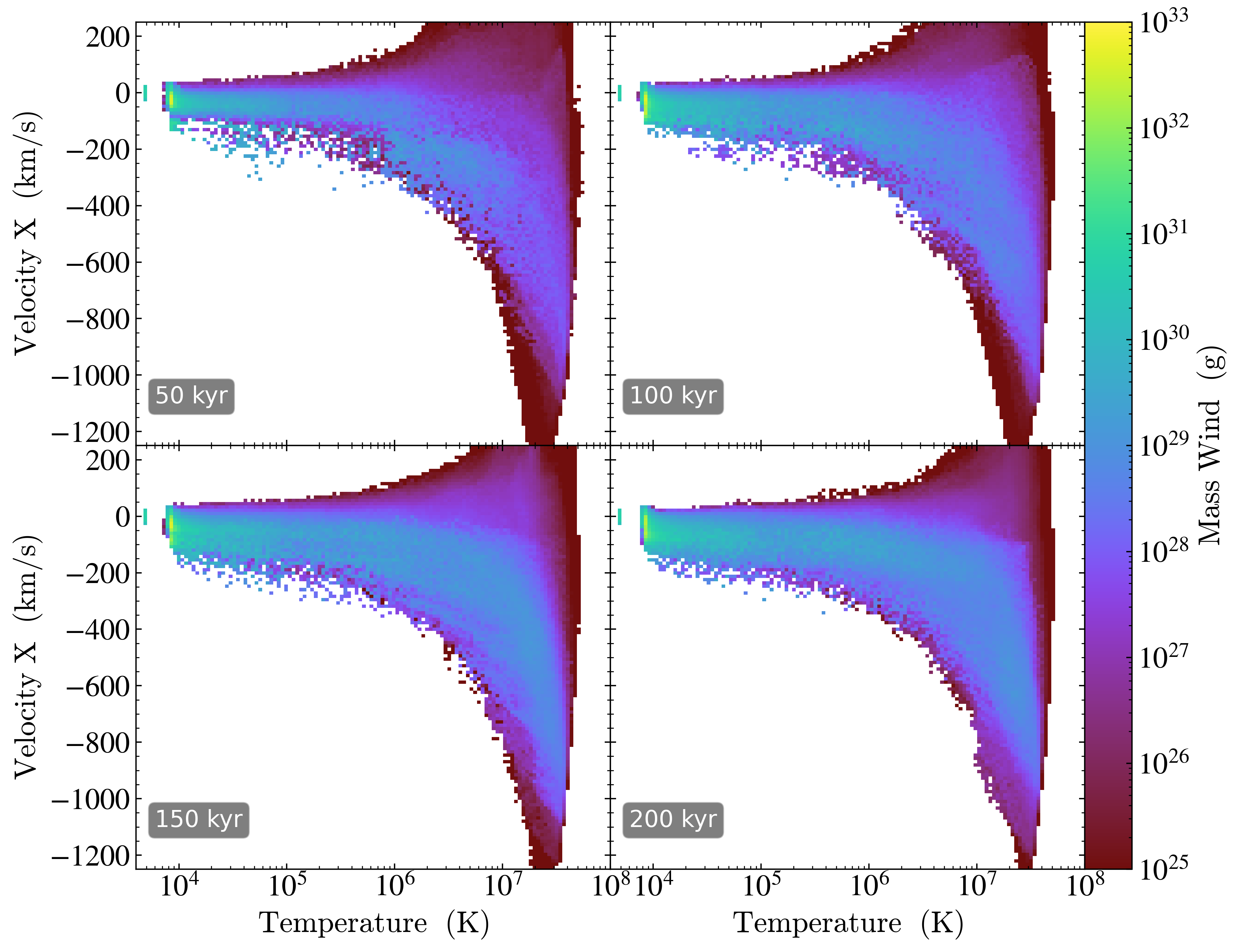}
    \caption{Same as Fig. \ref{fig:timeseries_vx_t} but for the lower-resolution model.}
    \label{fig:app_timeseries_vx_t}
\end{figure}

\begin{figure}
    \centering
    \includegraphics[width=1\linewidth]{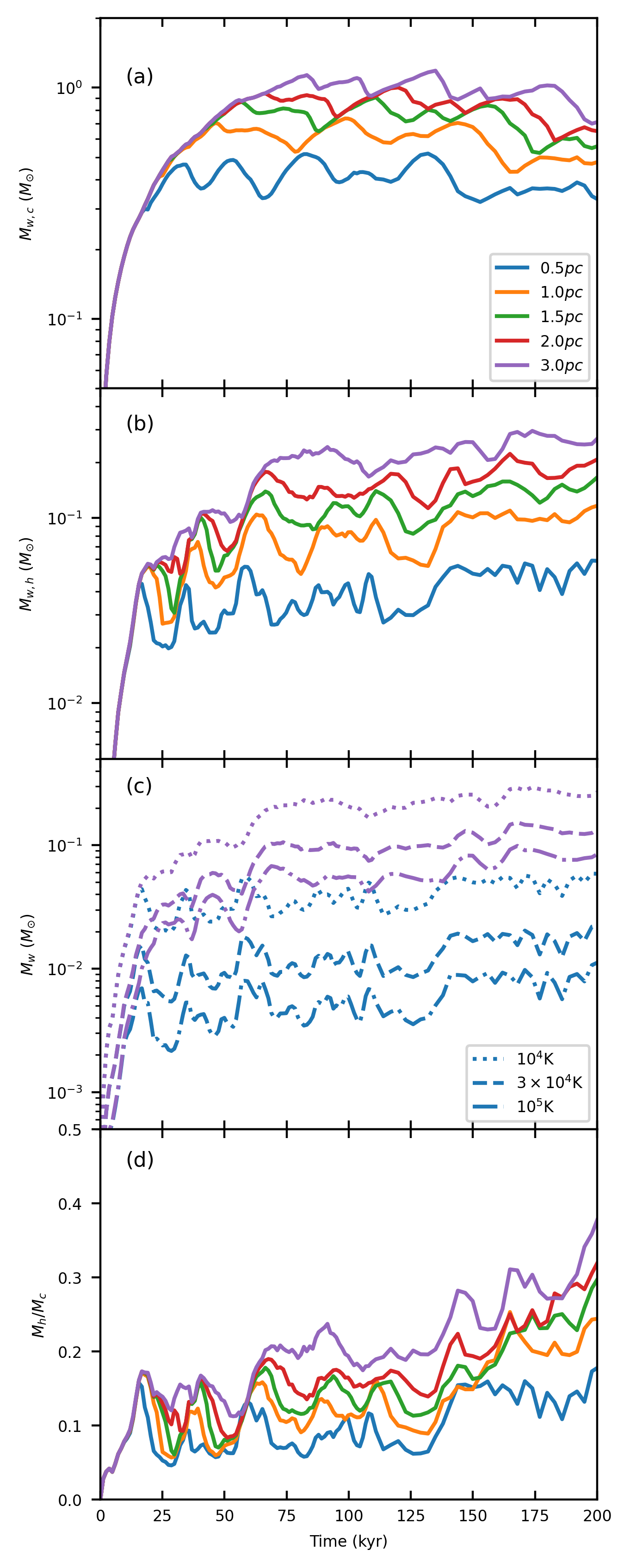}
    \caption{Same as Fig. \ref{fig:hot_cold_mass_t} but for the lower-resolution model.}
    \label{fig:app_hot_cold_mass_t}
\end{figure}

For a radius of 0.5 pc, both models have converged to similar values in all panels, and the differences likely originate from instabilities in the ablation flow. At 3.0 pc, the instabilities are more poorly resolved, reducing the efficiency of gas mixing. This suggests we do not have sufficient resolution at this radius, making our mass-loading estimate a lower limit. We present additional figures of our low resolution model as Figs. \ref{fig:app_end_xyxz_3panel}, \ref{fig:app_wind_ism_phase_end}, \ref{fig:app_timeseries_vx_t}, and \ref{fig:app_hot_cold_mass_t}.

Figure~\ref{fig:app_end_xyxz_3panel} shows that the downstream wake contains larger and more contiguous clumps that are more closely confined to the $x$-axis, when comparing this lower-resolution calculation with the higher-resolution simulation in Fig.~\ref{fig:end_xyxz_3panel}.
Despite this qualitative difference between the two resolutions, the phase diagrams (Figs.~\ref{fig:app_wind_ism_phase_end} and \ref{fig:app_timeseries_vx_t}) nevertheless appear quite similar between the two resolutions. 
The mass ratio of hot-to-cold gas in Fig.~\ref{fig:app_hot_cold_mass_t}(d) is significantly lower than in Fig.~\ref{fig:hot_cold_mass_t}.
This shows that dynamical mixing is not sufficiently well resolved at this lower numerical resolution,  and that higher resolution is required to demonstrate whether or not our results for mass loading have converged.
\end{document}